\documentclass[aps, pre, twocolumn,floatfix, superscriptaddress]{revtex4-2}

\usepackage{graphicx} 
\usepackage{float}
\usepackage[utf8]{inputenc}
\usepackage[T1]{fontenc}
\usepackage[english]{babel}
\usepackage{amsmath}
\usepackage{amssymb}
\usepackage{xcolor}
\usepackage{braket}
\usepackage{bm}
\usepackage{physics}
\usepackage[colorlinks=true, linkcolor=blue,
citecolor=red, 
urlcolor=blue, hypertexnames=true]{hyperref}

\usepackage{multibib}

\newcommand{\eref}[1]{Eq.~\eqref{#1}}

\newcommand{\fref}[1]{Fig.~\ref{#1}}
\newcommand{\Fref}[1]{Figure~\ref{#1}}

\newcommand{\p}{\partial}

\begin{document}

\title{Proxitaxis: An adaptive search strategy based on proximity and stochastic resetting}
\date{\today}

\author{Giuseppe Del Vecchio Del Vecchio}
\email{giuseppe.del-vecchio-del-vecchio@universite-paris-saclay.fr}
\affiliation{LPTMS, CNRS, Universit\'e Paris-Sud, Universit\'e Paris-Saclay, 91405 Orsay, France}

\author{Manas Kulkarni}
\email{manas.kulkarni@icts.res.in}
\affiliation{International Centre for Theoretical Sciences, Tata Institute of Fundamental Research,
Bangalore 560089, India}

\author{Satya N. Majumdar}
\email{satya.majumdar@universite-paris-saclay.fr}
\affiliation{LPTMS, CNRS, Universit\'e Paris-Sud, Universit\'e Paris-Saclay, 91405 Orsay, France}

\author{Sanjib Sabhapandit}
\email{sanjib@rri.res.in}
\affiliation{Raman Research Institute, Bangalore 560080, India}

\begin{abstract}
We introduce \emph{proxitaxis}, a simple search strategy where the searcher has only information about the distance from the target but not the direction. The strategy consists of three crucial components: (i) local adaptive moves with a distance-dependent  diffusion coefficient, (ii) intermittent long-range returns via stochastic resetting to a certain location  $\vec{R}_0$, and (iii) an inspection move where the searcher dynamically updates the resetting position $\vec{R}_0$. We compute analytically the capture probability of the target within this strategy and show that it can be maximized by an optimal choice of the control parameters of this strategy. Moreover, the optimal strategy undergoes multiple phase transitions as a function of the control parameters. These phase transitions are generic and occur in all dimensions. 
\end{abstract}

\maketitle

Search problems are ubiquitous in nature~\cite{BergPurcell1977, B1991,WH2004,MAEG09,F2015,GDM2024}. There are several examples where the searcher may have only partial information about the target. For instance, from some locally available cues, the searcher might be able to infer its distance from the target (mobile or immobile), but not the direction in which the target is located. A classic example of such search processes with incomplete information is the so-called ``infotaxis''~\cite{VVS07,LE22,CYBX2020,MM2010}, where unlike chemotaxis~\cite{J66, EL71,Payne1986, SW2012}, there is no strong concentration gradient to guide the searcher toward the target. This could be due to the existence of winds or turbulence that leads to sporadic or intermittent arrival of signals. Examples of such search processes with only partial information can be found in many contexts~\cite{dusenbery1992sensory, Murlis1992,Holdo2009, HM2012,berg2014coli,Fagan2017}. These include the search of a shipwreck deep inside the ocean using acoustic sensors, detection of a source of radioactivity or a gas leak, seismic exploration, and olfactory searches by animals. For example, when one searches for a radioactive source with a Geiger counter, it does not give any clue about the direction from which the radiation comes. In all these examples, the local sensor can give an estimate of its distance from the source, but not the direction in which the source is located. Thus, it is important and of practical relevance to engineer a search protocol where the searcher only has instantaneous information about the distance to the target.  

\begin{figure}[t!]    
\includegraphics[width=\linewidth]{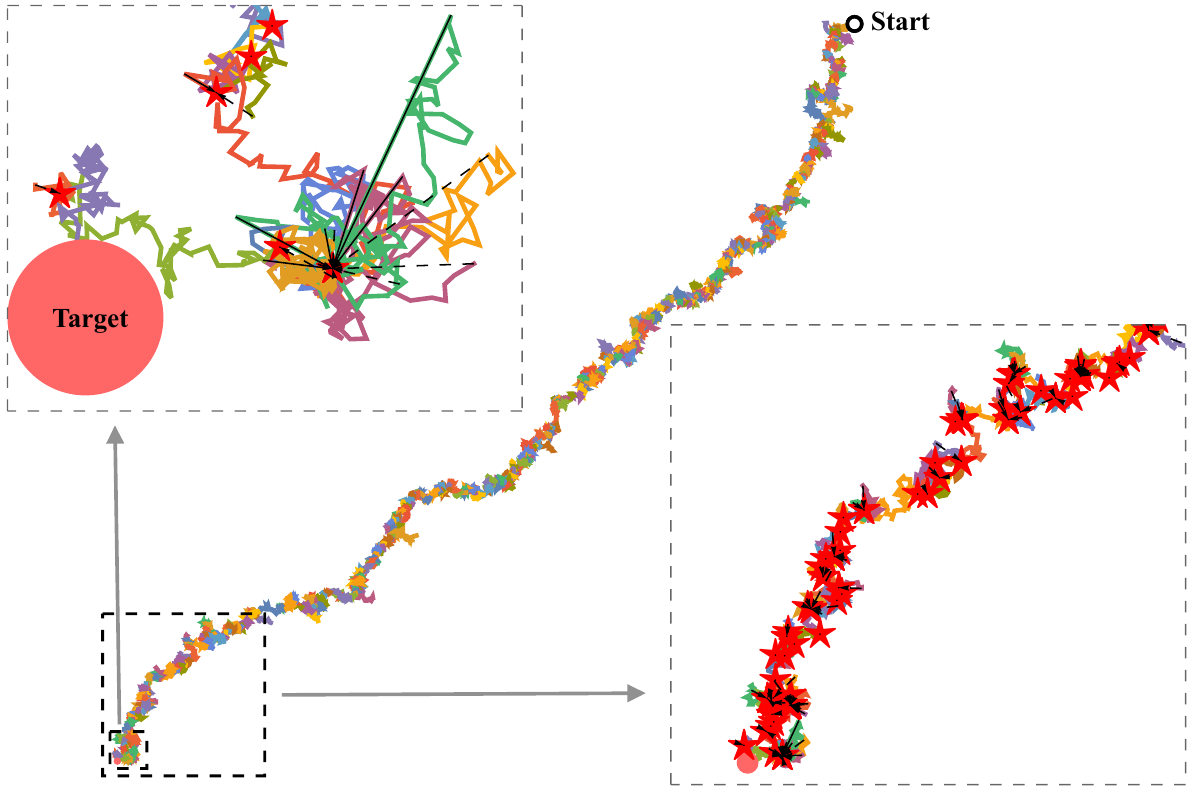}
    \caption{An \emph{optimal} trajectory generated by the \emph{proxitaxis} search strategy in two dimensions. The single target is shown by the red disk and the searcher's initial location is indicated by the black circle. The central figure shows the trajectory over a large length and time scale, illustrating that the strategy guides the searcher almost deterministically toward the target. However, on smaller scales, one sees fluctuations. For example, the top box zooms the trajectory close to the target, where the searcher becomes more active. In contrast, the right box zooms the trajectory at some intermediate scales where it is less active, i.e., more sluggish. See the main text for the details of the strategy. 
   } 
    \label{fig:traj}
\end{figure}

In this Letter, we propose a new search strategy that works efficiently with this limited information \emph{only} about distance from the source. We call this ``proxitaxis'' (inspired by the terminologies like chemotaxis and infotaxis), where ``proxi'' refers to proximity to the target and ``taxis'' to the movement of the searcher in response to stimuli.

One crucial ingredient of our \emph{proxitaxis} strategy is drawn from the so-called intermittent search protocol~\cite{benichou2005, Lomholt2008, benichou2007, BLMV2011}, where the searcher employs two types of moves intermittently: (a) local short-ranged stochastic jumps typically modeled by diffusion where the searcher tries to locate the target and (b) long-range moves during which it does not search but relocates to a different region of space. 
One popular model that incorporates these two intermittent moves, and yet is analytically tractable, is \emph{diffusion with stochastic resetting}~\cite{EM2011, EM2011a, Evans_2014, EMS20, Pal_2022, gupta2022, KS2024, Bressloff_2020a} that attracted much attention in recent times. In this model, the searcher performs Brownian diffusion (local moves) and, with a constant rate, resets instantaneously to its initial position (long-range moves) and restarts the search. The rationale is that when one restarts the search process, one may find a better pathway that leads to the target in a shorter time. In many model systems, it has been established that the stochastic resetting typically expedites the search of a target~\cite{EM2011, EM2011a, Evans_2013, Evans_2014, EMS20, Pal_2022, gupta2022, KS2024}, see however~\cite{R2016, PR2017, PP2019} for the conditions on the existence of a nonzero optimal resetting rate. Some of the analytical predictions have also been verified recently in experiments using colloidal particles in optical traps~\cite{besga2020, TPSRR2020, Faisant_2021}.

\begin{figure*}[t!]
    \centering   \includegraphics[width=\textwidth]{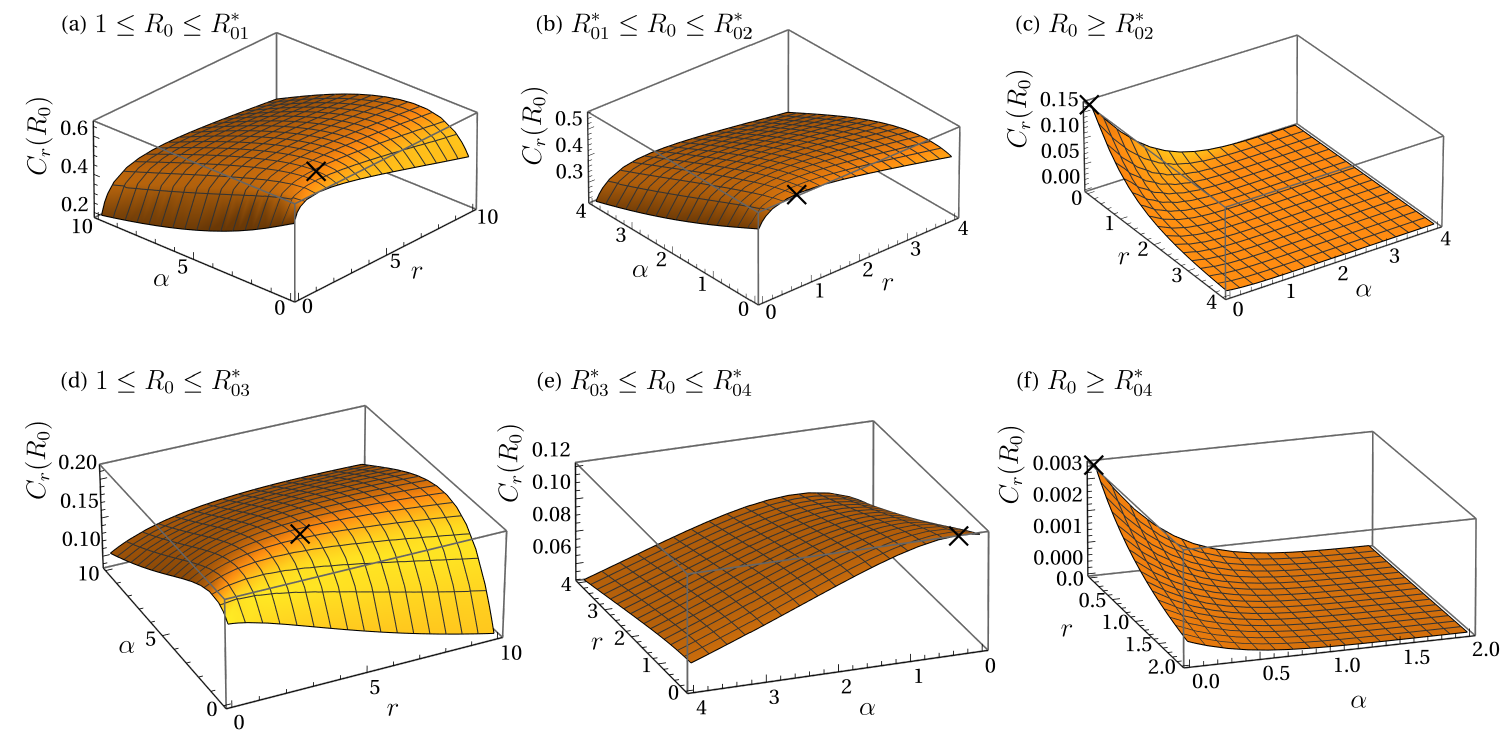}
    \caption{Three-dimensional plots of the capture probability $C(R_0)$ in~\eref{eq:capture1d} as a function of $(r,\alpha)$ for different values of $R_0$ in one dimension. The cross denotes the maximum  of $C(R_0)$ at the optimal values $(\overline{r}, \overline{\alpha})$. The top panels are for $b=0.2<b^*=1.194933\dots$ while the bottom panels are for $b=2>b^*$. (a) $R_0<R_{01}^*$ (b) $R_{01}^* <R_0<R_{02}^*$ (c) $R_0>R_{02}^*$ (d) $R_0<R_{03}^*$ (e) $R_{03}^* <R_0<R_{04}^*$ (f) $R_0>R_{04}^*$. Note that in general $R_{0i}^*$ for $i=2,3,4$ are nontrivial functions of $b$ (see~\fref{fig:optimal-alpha-r} for exact values corresponding to $b=0.2$ and $b=2$). 
 }
    \label{fig:optimalCP}
\end{figure*}

In typical diffusive search processes the diffusion coefficient is assumed to be independent of space~\cite{BLMV2011}.  However, in many situations the diffusion coefficient may depend on space, e.g., it can change in response to the lack or the abundance of resources available during foraging~\cite{VISWANATHAN2008133, Chupeau2017, hein2016}.  In some cases, animals are known to modify their movements when they get closer to their targets~\cite{ Viswanathan_daLuz_Raposo_Stanley_2011, Jashnsaz_2017}. Far from the target, the movement is typically slow or sluggish with a lower diffusion coefficient, allowing a broader local exploration. However, when the animal gets closer to the target, it can detect cues such as smell or sound, which may trigger faster and more active movements with an effective diffusion coefficient which is larger, allowing it to scan for the target more intensively.
Sperm cells are known to undergo  hyperactivation that changes their motility state as they approach the egg cell~\cite{SHS2003, ZB2025}.  
Another illustrative example of changing diffusivity can be found in searchers performing active Brownian motion, where the particle moves ballistically with speed $v_0$, but whose direction undergoes a rotational diffusion with diffusion coefficient $D_\phi$. At late times, the process becomes diffusive with an effective diffusion coefficient~\cite{Basu2024, BMRS2018} $D = v_0^2/(2D_\phi)$. 
It is reasonable to assume that when an active searcher senses that the target is near, it
increases its speed $v_0$ or reduces its rotational diffusivity
$D_\phi$ to stay in course, thereby increasing its effective diffusion coefficient $D$.
However, just a space-dependent diffusion coefficient is not enough to optimize the search of the target. This is because some trajectories may take the searcher away from the target via diffusion. A simple and inexpensive strategy to eliminate such rogue trajectories that does not require keeping memory of the history is to stochastically reset the walker to its initial position~\cite{EM2011, EM2011a}. In our \emph{proxitaxis} search strategy, we incorporate these two important ingredients: an effective space-dependent diffusion coefficient that increases as the searcher nears the target and stochastic resetting to its initial position.
We note that any search process will, of course, involve some microscopic length and time scale describing the process. Here, for simplicity, we will express all length and time scales in units of these microscopic length and time units, rendering them dimensionless.

These two components, namely the distance-dependent local exploration and stochastic resetting to the initial location $\vec{R}_0$, are, however, still not enough to ensure an efficient search since the walker has at its disposal the information about its instantaneous distance from the target. It makes no sense to reset to $\vec{R}_0$ if its current distance  $R(t)<R_0=\|\vec{R}_0\|$. Therefore, the searcher should ideally compare its current distance $R(t)$ to $R_0$ and should reset to $\vec{R}_0$ only if $R(t)>R_0$. Moreover, to make the search more efficient, the walker should also dynamically update its resetting position from $\vec{R}_0$ to $\vec{R}(t)$ if $R(t)<R_0$. It is, however, highly costly and hence inefficient to do this comparison at every step. It is natural to assume that there is a typical timescale associated with this inspection process, which depends on the animal's processing ability. Here, we simply assume that this inspection happens intermittently at Poisson-distributed times with a rate $b$.

In this Letter, we propose \emph{proxitaxis} as a possible new search strategy based on
these three aspects of a realistic search with only information about the distance from the
target. The \emph{proxitais} protocol is essentially a recursive search strategy with
memory that proceeds in three steps:

(1) One first chooses a random interval $\tau_1$ from $p(\tau)=b \,e^{-b \tau}$. Let $\vec R_0^{(1)}$
be the position at the beginning of this interval. During the interval, the walker performs diffusion
with diffusion coefficient $D(R)$ that depends on the instantaneous distance $R$ of the walker from the
target and resets to its initial position $\vec R_0^{(1)}$ with rate $r$. Let $\vec R (\tau_1)$ denote
the position of the walker at the end of the interval.

(2) The walker inspects its current position $\vec R (\tau_1)$ at the end of the first interval and compares it with the position
$\vec R_0^{(1)}$ at the beginning of this interval. If the current distance $||\vec R (\tau_1)||$ is bigger than 
the initial distance, then the ``new reset'' position $\vec R_0^{(2)}$ for the next interval is set to be equal to $\vec R_0^{(1)}$. In
the opposite case, the ``new reset'' position   $\vec R_0^{(2)}$ is chosen to be the position at the end of the first interval, i.e.,
$\vec R_0^{(2)}=\vec R (\tau_1)$.

(3) Once the initial position $\vec R_0^{(2)}$ is chosen, the walker chooses another interval $\tau_2$ from $p(\tau)= be^{-b\tau}$
and again carries out the diffusion with diffusion coefficient  $D(R)$ and resetting stochastically to $\vec R_0^{(2)}$ with rate $r$.

These three steps are repeated with independently drawn random intervals $\{\tau_i\}$ and a sequence of resetting positions $\{R_0^{(i)}\}$, till the target is found. The inspection step~2 is indeed an important
component of the \emph{proxitaxis} strategy, in addition to space-dependent diffusion and resetting, whereby the resetting location
is dynamically updated. This creates an overall drift toward
the target, making the strategy highly effective~[see~\fref{fig:traj}, where the sequence of
$R_0^{(i)}$'s are marked by the red stars]. 
Given the set of intervals $\{\tau_1,\tau_2,\tau_3, \dotsc\}$, one can choose the diffusion coefficient $D(R)$ and the resetting rate $r$ in order to maximize the capture probability independently within each interval $\tau_i$. In our proxitaxis strategy, an
efficient searcher does this optimization independently within each interval. Solving analytically a simple $d$-dimensional model, we show
that the capture probability within each interval $\tau_i$ can indeed be optimized by a
suitable choice of the control parameters $\{D_i(R),r_i\}$ in any dimension. With this
choice of the optimal control parameters, the \emph{proxitaxis} search strategy can then be
made efficient in all dimensions. Note that although $b$ is also a parameter of the model, it characterizes an intrinsic property of the searcher and is kept fixed during the optimization with respect to $\{D_i(R), r_i\}$.

Since we optimize the parameters $\{D_i(R),r_i\}$ independently within each time interval $\tau_i$, to find the optimal parameters, one can just focus on a single interval. Therefore, we will drop the index $i$ labeling the intervals in what follows. We consider a searcher in $d$-dimensions in the presence of a spherical target of radius $\epsilon>0$. Since only the relative position between the searcher and the target matters, without loss of generality, we place the target at the origin. Let $\vec{R}_0$ be the position of the searcher at the beginning of the interval.
Given this fixed interval, we have two dimensionless spatial parameters ($\epsilon$ denoting the target size and $\vec{R}_0$ denoting the initial distance) and two dimensionless temporal parameters ($b^{-1}$ denoting the lifetime of the interval and $r^{-1}$ denoting the mean time between two resettings). Given these parameters,  
the dimensionless position $\vec{R}(t)$ of the searcher evolves via the following dynamics. In a small time interval $[t,t+\Delta t]$, it resets to $\vec{R}_0$ with probability $r\Delta t$ and with the complementary probability $1-r \Delta t$, it diffuses with the diffusion coefficient $D(R(t))$ that depends only on the distance $R(t) = ||\vec{R}(t)||$ of the instantaneous position of the searcher from the target. The search process terminates when the searcher hits the boundary of the target, i.e., when $R(t) = \epsilon$.  However, within the interval $\tau$, the searcher may or may not find the target. In order to characterize the efficiency of the strategy, we maximize the capture probability $C(\vec{R}_0)$ defined as the probability that the target is captured within the lifetime of this strategy.

To compute $C(\vec{R}_0)$ we proceed as follows. Let 
$Q_r(\vec{R}_0,t)$ denote the survival probability of the
target up to time $t$, which is the probability that the target is not yet detected up to $t$.
Then $[-\partial_t Q_r(\vec{R}_0,t)]$
is the first-passage probability, i.e., the capture probability at time $t$.
The probability of survival of the strategy up to time $t$ is simply $q(t)= \int_t^\infty p(\tau)\, d\tau=\int_t^{\infty} b\, e^{-b \tau}\, d\tau=e^{-b t}$. 
Hence, the capture probability within this strategy is  given by
\begin{equation}
C(\vec{R}_0)=  \int_0^{\infty} dt\, \left[-\partial_t Q_r(\vec{R}_0,t)\right]\, e^{-b\, t}\, .
\label{capture.1m}
\end{equation}
Performing integration by parts and using $Q_r(\vec{R}_0,0)=1$ one gets
\begin{equation}
C(\vec{R}_0)=  \left[1- b\, {\tilde Q}_r(\vec{R}_0, b)\right]\, ,
\label{capture.2m}
\end{equation}
where ${\tilde Q}_r(\vec{R}_0,s)= \int_0^{\infty} Q_r(\vec{R}_0,t)\, e^{-s\, t}\, dt$ is the Laplace transform of the
survival probability. The survival probability (rather its Laplace transform) $\tilde Q_r(\vec{R}_0,s)$ in the presence of resetting can be related to the same quantity $\tilde Q_0(\vec{R}_0,s)$ in the absence of resetting via the well-known renewal relation~\cite{KMSS14, R2016, EMS20}
\begin{equation}
{\tilde Q}_r(\vec{R}_0,s)= \frac{{\tilde Q}_0(\vec{R}_0, r+s)}{1- r\, {\tilde Q}_0(\vec{R}_0, r+s)}\, .
\label{renewal.1m}
\end{equation} 
Substituting \eref{renewal.1m} into \eref{capture.2m} gives
\begin{equation}
C(\vec{R}_0)= \left[ \frac{1- (r+b)\, {\tilde Q}_0(\vec{R}_0, r+b)}{1- r\, {\tilde Q}_0 (\vec{R}_0, r+b)}\right]\, .
\label{capture.3m}
\end{equation} 
The survival probability $Q_0(\vec{R}_0,t)$, in the absence of resetting, satisfies the backward Fokker-Planck equation (see Supplemental Materials (SM)~\cite{supp})
\begin{equation}
\frac{\p Q_0(\vec{R}_0, t)}{\p t} = D(R_0)\nabla^2 Q_0(\vec{R}_0,t)
\label{eq:bwfp}
\end{equation}
with the initial condition $Q_0(\vec{R}_0,t=0)=1$ for all $R_0>\epsilon$. The boundary condition is absorbing at $R_0=\epsilon$, i.e.,  $Q_0(R_0=\epsilon, t)=0$ and also far from the target $Q_0(R_0\to + \infty, t)$ remains finite.  Since $D(R_0)$ depends only on the radial distance $R_0$, the survival probability $Q_0(\vec{R}_0, t)=Q_0(R_0,t)$ also has spherical symmetry. Consequently, Eq. \eqref{eq:bwfp} becomes
\begin{equation}\label{eq:bw_survival_d}
     \frac{\p Q_0}{\p t} = D(R_0) \Bigg[\frac{\p^2 Q_0}{\p R_0^2}   + \frac{d-1}{R_0}\frac{\p Q_0}{\p R_0}\Bigg]\, .
\end{equation}
From this backward Fokker-Planck equation, one can read off the associated Langevin equation for the evolution of the radial coordinate $R(t)$
\begin{equation}\label{eq:langevin_d}
    \frac{\dd R}{\dd t} = \frac{d-1}{R}D(R)+ \sqrt{2 D(R)} \eta(t)
\end{equation}
where $\eta(t)$ is Gaussian white noise with zero mean and correlator
$\langle \eta(t)\eta(t')\rangle = \delta(t-t')$, and the noise term has to be interpreted in the It\^o sense since $D(R(t)) \equiv D(R)$ depends only on the instantaneous position at time $t$. When $D(R)$ is a constant independent of $R$, Eq.~\eqref{eq:langevin_d} just represents the well-known Bessel process~\cite{revuz2013}. Such Langevin dynamics with multiplicative noise (or associated Fokker-Planck equations) have also been widely studied in the context of the dynamics of subdiffusion in heterogeneous media~\cite{KL2003, CCM2013,   LLGER2022, ZAEM23, SCT2023, SCT2023PRE,  MC2024, Del_Vecchio_Del_Vecchio_2025, EISINGER1986987, SAXTON1987989, SAXTON1989615, S1993, Nicolau2007, masuhara2007general, Singh_2020}.

To make further analytical progress, we need to specify a form of $D(R)$ in~\eref{eq:bw_survival_d}.
In principle, one can study any $D(R)$ that decreases with $R$. However, a natural choice is a scale-free power law $D(R) = R^{-\alpha}$ for all $R>0$. This form of $D(R)$, as we will see, makes the analytical calculations possible. In more realistic models, the power law form does not hold all the way up to $R=0$ and usually there is a cutoff, say at $R=1$ (we recall here that $R$ is dimensionless), which regularizes the unphysical divergence as $R\to 0$. However, in the presence of the cutoff at $R=1$, finding an analytical solution becomes very hard. Hence, we will assume that the power law form of $D(r)$ holds all the way up to $R=0$, with the caveat that this solution is not reliable for $R<1$. 
Incidentally, this scale-free choice of $D(R)$ has been studied in the context of transport in one-dimensional inhomogeneous systems~\cite{ZAEM23}, and several statistical properties have been computed~\cite{Del_Vecchio_Del_Vecchio_2025}.
For this choice of $D(R)$, the survival probability $Q_0(R_0,t)$ can be obtained by solving~\eref{eq:bw_survival_d}. Taking the Laplace transform of~\eref{eq:bw_survival_d} with respect to $t$, it is easy to show that (see SM~\cite{supp}),
\begin{equation}\label{eq:surv_nr}
    \tilde Q_0(R_0, s) = \frac{1}{s}\left[1 -  \left(\frac{R_0}{\epsilon}\right)^{\frac{2-d}{2}} \frac{K_{(d-2)\mu}\left(2\mu \sqrt{s}R_0^{\frac{1}{2\mu}}\right)}{K_{(d-2)\mu}\left(2\mu \sqrt{s}\epsilon^{\frac{1}{2\mu}}\right)}\right]
\end{equation}
where $\mu = 1/(\alpha + 2)$, $K_\nu(x)$ is the modified Bessel function of the second kind, and we recall that $\epsilon$ is the size of the target. Substituting~\eref{eq:surv_nr} in~\eref{capture.3m} gives the exact capture probability $C(\vec{R}_0)\equiv C(R_0)$ as
\begin{equation}
 C(R_0)=\frac{(b+r)}{r + b\,  \left(\frac{R_0}{\epsilon}\right)^{\frac{d-2}{2}}\,
\frac{ K_{(d-2)\mu}\left(2\mu\, \sqrt{r+b}\, \epsilon^{1/{2\mu}}\right)}
{ K_{(d-2)\mu}\left(2\mu\, \sqrt{r+b}\, R_0^{1/{2\mu}}\right)} }\, .
\label{capture_final1}
\end{equation}
This is one of our principal analytical results. As in the standard Brownian diffusion (the case $\alpha=0$), it
turns out for any $\alpha>0$, the size $\epsilon$ of the target can be made as small as possible
in $d<2$, while for $d\ge 2$, one needs to keep a
finite nonzero $\epsilon$. This is because the
trajectory will always miss, with probability $1$,
a point target for $d\ge 2$~\cite{Redner_2001, Bray_2013}. For example, in $d=1$ upon taking the $\epsilon\to0$ limit, Eq.~\eqref{capture_final1} simplifies to
\begin{equation}
\label{eq:capture1d}
    C(R_0)=\frac{b+r}{r+ \frac{b\,\Gamma(\mu)}{2 \sqrt{R_0}\,\mu^\mu \, (r+b)^{\frac{1}{2\mu}}\, K_\mu\left(2\mu\, \sqrt{r+b}\, R_0^{\frac{1}{2\mu}}\right)}}
\end{equation}
where we recall that $\mu=1/(2+\alpha)$. In \fref{fig:optimalCP}, we plot $C(R_0)$ in~\eref{eq:capture1d} in the $(r,\alpha)$ plane. \Fref{fig:optimalCP} clearly demonstrates that there is a unique global maximum of $C(R_0)$ at an optimal value $(\overline{r},\overline{\alpha})$. The same feature holds in higher dimensions. We show this explicitly in $d=2$ and $d=3$ in SM~\cite{supp}. Figure~\ref{fig:traj} displays an optimal trajectory for $d=2$ of the search strategy where within each lifetime the parameters $(r,\alpha)$ take their optimal values $(\bar r, \bar \alpha)$. See SM~\cite{supp} for a plot of a typical trajectory with sub-optimal parameters.

Hence, for fixed $R_0$, $\epsilon$, and $b$ we can optimize $C(R_0)$ in Eq.~\eqref{capture_final1} in the $(r,\alpha)$ plane and find the optimal parameters $(\overline{r}, \overline{\alpha})$. It is simpler to carry out this optimization in $d=1$ (and $\epsilon\to 0$ limit) where we can use Eq.~\eqref{eq:capture1d}. Our analysis demonstrates rather interesting and rich phase transitions in these optimal values $(\overline{r}, \overline{\alpha})$ as one tunes $R_0$ and $b$. We now summarize these results.
\begin{figure}[t!]
    \centering
\includegraphics[width=\linewidth]{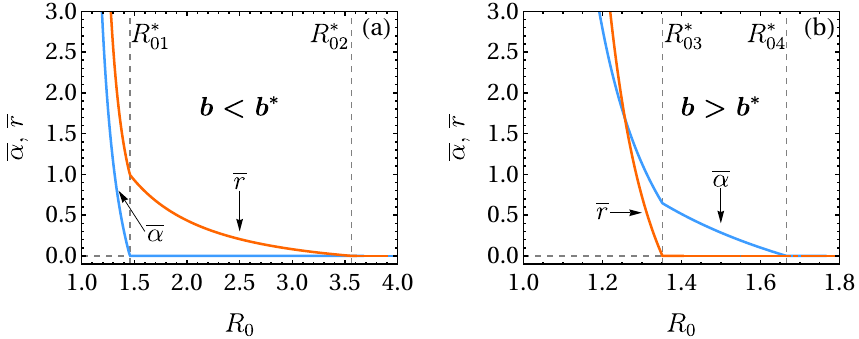}
    \caption{Plot of $\overline{\alpha}$ and $\overline{r}$, as a function of $R_0$ for (a) $b<b^*=1.19494\dots$ and (b) $b>b^*$, in one dimension, obtained by maximizing~\eref{eq:capture1d}. 
    For both cases, $\overline{r}$ and $\overline{\alpha}$ are nonanalytic at some critical values of $R_0$ (see SM~\cite{supp}) marked by the vertical dash lines. For panel (a) we choose $b=0.2<b^*$, for which the critical values are at $R_{01}^*=1.4578\dots$ and $R_{02}^*=3.5634\dots$.
    For (b), we choose
    $b=2 > b^*$, for which the critical values of $R_0$ are at $R_{03}^*=1.35216\dots$ and $R_{04}^*=1.6661\dots$. 
    } 
    \label{fig:optimal-alpha-r}
\end{figure}
The optimal values $(\overline{r}, \overline{\alpha})$, as functions of $R_0$ and for fixed $b$, undergo a phase transition at two critical values of $R_0$. However, the nature of the phases depends on $b$. It turns out that there is a threshold value $b^*$, across which two different scenarios emerge: 

(1) For $b<b^*$, we denote the two critical values as $R_{01}^*$ and $R_{02}^*$. For $R_0>R_{02}^*$, the
optimal strategy corresponds to standard diffusion \emph{without} resetting, i.e.,  $\overline{r}=0$ and $\overline{\alpha} = 0$. For $R_{01}^* < R_0 < R_{02}^*$, the optimal strategy is standard diffusion \emph{with} stochastic resetting~\cite{EM2011, EM2011a}, i.e., $\overline{r}> 0$ and $\overline{\alpha} = 0$.
Finally, for $R_0<R_{01}^*$, the optimal strategy has both $\overline{r}>0$ and $\overline{\alpha}>0$.

(2) For $b>b^*$, we denote the two critical values of $R_0$ by $R_{03}^*$ and $R_{04}^*$. In this case, again $\overline{r}=0$ and $\overline{\alpha} = 0$ for $R_0>R_{04}^*$  and both $\overline{r}>0$ and $\overline{\alpha}>0$ for $R_0<R_{03}^*$. In contrast to the $b<b^*$ case, here, for $R_{03}^* < R_0 <R_{04}^*$, we have $\overline{r}(R_0)= 0$ and $\overline{\alpha} >  0$.

This is illustrated in~\fref{fig:optimal-alpha-r} for $d=1$. However, these transitions turn out to be robust and generic and are also present in higher dimensions. We demonstrate this for $d=2$  and $d=3$ (see SM~\cite{supp}).  This nontrivial phase transition in the optimal parameters as functions of the initial distance $R_0$ (for a fixed target size $\epsilon$) is our second main result. We also note that, as $R_0\to 1^+$, for any $b$, both the optimal parameters diverge as (see SM~\cite{supp}) 
\begin{equation}\label{eq:power_law_divergence}
    \overline{r}\approx \frac{9}{16 \,e \, (R_0 - 1)^2}~ \quad\text{and}\quad \overline{\alpha} \approx \frac{1}{(R_0-1)} 
\end{equation}
where $e=\exp(1)$. It is interesting that \eref{eq:power_law_divergence} is universal at the leading order for all dimensions, and it is independent of the target size $\epsilon$.
For $R_0 \le 1$, the optimal pair of values freezes to $\overline{r}=+\infty$ and $\overline{\alpha}=+\infty$, and one can show that the optimized capture probability converges to unity. 
The origin of this unphysical divergence as $R_0 \to 1^+$ can be traced back to the fact that the analytical solution uses the scale-free form $D(R)=R^{-\alpha}$ all the way up to $R=0$. As mentioned earlier, in reality, there is always a short-distance regularization of $D(R)$ that cuts off this divergence. Hence, our analytical solution provides a good description only for $R>1$. Nonetheless, the phase transitions in the optimal parameters $(\bar r, \bar \alpha)$ found in our analytical solution are expected to be robust and independent of the precise regularization used for $D(R)$ for small $R$.

To summarize,  we have studied search processes in situations where only partial information, namely, the distance to the target but not the direction, is available to the searcher. 
We have introduced the \emph{proxitaxis} search strategy and demonstrated that it can be made efficient in all dimensions by choosing an appropriate set of optimal parameters of this strategy. Our results demonstrate a very rich behavior, exhibiting multiple phase transitions in the optimal parameters in all dimensions. 

\emph{Proxitaxis is a simple}, analytically tractable search strategy, which can be optimized explicitly with respect to certain parameters. There exist very few realistic search strategies for which this can be done. Ours is thus a rare analytically optimizable search strategy. The question of whether real animals follow this strategy is obviously legitimate and we are not claiming that this is the most efficient search strategy. Rather, our goal is to provide a benchmark strategy which is totally controllable analytically and helps us understand the microscopic optimal moves much better. 
 We hope that our results will stimulate interest in comparing real-life animal movement data with our predictions. Furthermore, even if natural animals do not exactly follow this strategy, it can still be very useful for designing artificial search robots operating in situations where there is no orientational information available. A detailed comparison between different strategies using different rules than ours is an interesting question, but it requires a careful selection of meaningful metrics. In this paper we have not tried to make such a comparison and leave it as an interesting future direction.

Furthermore, there are several directions in which the  \emph{proxitaxis} strategy can be generalized. For example, one might consider multiple targets~\cite{bressloff_2020b, Calvert2021} and/or multiple searchers~\cite{Bhat_2016, biroli_2023}. Moreover,  instead of pure diffusion as the basic dynamics of the searcher, one could consider active Brownian motion, e.g., in two dimensions. This introduces an additional parameter representing the activity rate, which can be used for further optimization. Another interesting generalization would be to the case where some partial information about the direction of the target is available to the searcher. How do the optimal parameters depend on this partial information? In our model, we assumed that the stochastic resetting is memoryless, in the sense that the walker always resets to its initial position. It would be interesting to incorporate the effect of memory, where the walker remembers the positions it has visited before. Furthermore, as in chemotaxis, it would be interesting to introduce a drift in addition to the diffusion that depends on the local concentration gradient of the resources. The explicit results presented here with a simple proxitaxis model thus clearly illustrate that there are several open avenues for future research.

{\em Acknowledgements.} We thank D. Boyer, L. Giuggioli, and V. Krishnamurthy for providing us with interesting
references. We acknowledge support from the Science and Engineering Research Board (SERB, Government of India), under the VAJRA faculty scheme (VJR/2017/000110 and VJR/2019/00007). M. K. acknowledges support from the Department of Atomic Energy, Government of India, under project no. RTI4001.  G. D. V. D. V. and S. N. M. acknowledge support from ANR Grant No. ANR-23-CE30-0020-01 EDIPS.

\bibliographystyle{apsrev4-2-titles}
\bibliography{references}

\end{document}


\title{Proxitaxis: an adaptive search strategy based on proximity and stochastic resetting}

\date{\today}

\author{Giuseppe Del Vecchio Del Vecchio}
\email{giuseppe.del-vecchio-del-vecchio@universite-paris-saclay.fr}
\affiliation{LPTMS, CNRS, Universit\'e Paris-Sud, Universit\'e Paris-Saclay, 91405 Orsay, France}

\author{Manas Kulkarni}
\email{manas.kulkarni@icts.res.in}
\affiliation{International Centre for Theoretical Sciences, Tata Institute of Fundamental Research,
Bangalore 560089, India}

\author{Satya N. Majumdar}
\email{satya.majumdar@universite-paris-saclay.fr}
\affiliation{LPTMS, CNRS, Universit\'e Paris-Sud, Universit\'e Paris-Saclay, 91405 Orsay, France}

\author{Sanjib Sabhapandit}
\email{sanjib@rri.res.in}
\affiliation{Raman Research Institute, Bangalore 560080, India}

\begin{abstract}
    In this Supplemental Material, we provide the detailed derivations of some of the results presented in the main text.
\end{abstract}

\maketitle

\thispagestyle{fancy}


\tableofcontents

\section{Details for the derivation of survival probability}
\label{sec:sp}
In this section, we discuss how to derive Eq.~(8) of the main text. This quantity has been studied for $d=\alpha=1$ in \cite{ZAEM23} and generalized to all $\alpha\in[0,+\infty)$ in \cite{Del_Vecchio_Del_Vecchio_2025} via a scaling argument. Here, we further generalize those results to arbitrary $\alpha\in[0,+\infty]$ and arbitrary dimensions $d$. The use of the backward Fokker-Planck approach is motivated by the fact that it directly gives access to $\tilde Q_0(R_0,s)$, i.e., the Laplace transform, which is the main ingredient for the capture probability given in (4) of the main text.

We start from the Langevin equation Eq. (7) of the main text, which reads
\begin{equation}\label{eq:langeving_d_dim}
    \frac{\dd R}{\dd t} = \frac{d-1}{R}D(R)+ \sqrt{2 D(R)} \eta(t)\, ,
\end{equation}
where the multiplicative white noise term is interpreted in the It\^o sense. It is well known that the forward Fokker-Planck equation for the position probability distribution $P_0(R,t)$, in the absence of resetting, reads \cite{ZAEM23, SCT2023, SCT2023PRE,  MC2024, Del_Vecchio_Del_Vecchio_2025}
\begin{equation}\label{eq:forward_fp_supp}
    \frac{\p P_0}{\p t} = \frac{\p^2}{\p r^2}\left[D(R)P_0\right]  - \frac{\p}{\p r}\left[\frac{d-1}{R}D(R)P_0\right]\,.
\end{equation}
In contrast, the survival probability $Q_0(R,t)$, where $R_0$ is the starting position, satisfies the backward Fokker-Planck equation (the adjoint of Eq.~\eqref{eq:forward_fp_supp})
\begin{equation}
    \label{eq:back_fpe}
      \frac{\p Q_0}{\p t} =D(R_0) \Bigg[\frac{\p^2 Q_0}{\p R_0^2}   + \frac{d-1}{R_0}\frac{\p Q_0}{\p R_0}\Bigg]\, ,
\end{equation}
valid for $R_0>\epsilon$. It satisfies the boundary
\begin{equation}\label{eq:bc_backward_equation}
    Q_0(R_0=\epsilon,t) = 0\quad \text{and} \quad Q(R_0\to +\infty,t)=1\,.
\end{equation} 
The first condition comes from the fact that when the boundary $R_0=\epsilon$ is absorbing, i.e., if the particle starts from this boundary, it does not survive up to any finite time $t$. The second boundary condition comes from the fact that if the particle starts very far from the target, it definitely survives up to any finite time $t$ with probability $1$. Eq.~\eqref{eq:bc_backward_equation} starts from the initial condition
\begin{equation}\label{eq:ic_backward_equation}
      Q_0(R_0,0) = 1\quad \text{for all}\quad R_0>\epsilon\,.
 \end{equation}
 Here $0<\epsilon<R_0$ is the radius of the finite-size target.
To proceed, it is convenient to define the Laplace transform with respect to $t$ as
\begin{equation}
     \tilde Q_0(R_0,s) = \int_0^{+\infty} d \tau \, e^{-s \tau} Q_0(R_0,\tau)\, .
 \end{equation}
Taking the Laplace transform of Eq.~\eqref{eq:bc_backward_equation} and with the choice $D(R)=R^{-\alpha}$ one gets
 \begin{equation}
     - 1 + s\,\tilde Q_0 = \frac{1}{ R_0^\alpha}\frac{\p^2 \tilde Q_0}{\p R_0^2} + \frac{d-1}{R_0^{\alpha + 1}}\frac{\p \tilde Q_0}{\p R_0}\, . 
 \end{equation}
Shifting $\tilde Q_0$ by $1/s$ and solving the resulting homogeneous equation, one gets
\begin{equation}\label{eq:Qtilded}
     \tilde Q_0(R_0,s) = A\,R_0^{\frac{2-d}{2}}K_{\mu(d-2)}\left(2\mu s^{\frac{1}{2}}R_0^{\frac{1}{2\mu}}\right) + B\,R_0^{\frac{2-d}{2}}I_{\mu(d-2)}\left(2\mu s^{\frac{1}{2}}R_0^{\frac{1}{2\mu}}\right) + \frac{1}{s}\, ,
\end{equation}
where $A$ and $B$ are constants and $I_\mu(x)$ and $K_\mu(x)$ are modified Bessel functions of the first and second kind, respectively. From the known asymptotic behaviors of the modified Bessel functions \cite{abramowitz1965handbook} 
\begin{equation}\label{eq:large_x_bessel}
    K_\mu(x) \sim \sqrt{\pi / (2x)}\,e^{-x} \quad\text{and} \quad I_\nu(x) \sim \sqrt{\pi / (2x)}\,e^{x}\quad \text{as}\quad x\to+\infty \, ,
\end{equation}
we see that the first condition in \eref{eq:bc_backward_equation} imposes $B =0$ because that term is divergent. To find the constant $A$, we impose the second boundary condition $ \tilde Q(\epsilon, s) = 0$ in Eq. \eqref{eq:bc_backward_equation}. This gives
\begin{equation}
\label{eq:Acond}
    A = - \frac{1}{s}\frac{1}{\epsilon^{\frac{2-d}{2}}K_{\mu(d-2)}\left(2\mu s^{\frac{1}{2}}\epsilon^{\frac{1}{2\mu}}\right)}\quad .
\end{equation}
Substituting $A$ from~\eref{eq:Acond} and $B=0$ in \eref{eq:Qtilded} gives 
\begin{equation}\label{eq:surv_nr_supp}
    \tilde Q_0(R_0, s) = \frac{1}{s}\left[1 -  \left(\frac{R_0}{\epsilon}\right)^{\frac{2-d}{2}} \frac{K_{(d-2)\mu}\left(2\mu \sqrt{s}R_0^{\frac{1}{2\mu}}\right)}{K_{(d-2)\mu}\left(2\mu \sqrt{s}\epsilon^{\frac{1}{2\mu}}\right)}\right]\, ,
\end{equation}
which is Eq.~(8) in the main text.

\section{Relation between the capture probability and the mean first passage time}
\label{sec:rel}
In this section, we establish an exact relation, in the presence of resetting, between the capture probability $C(R_0)$  and the mean first passage time (MFPT), denoted $T_r(R_0)$. Let $F_r(R_0,t)$ denote the first passage probability density to the target for a particle starting at $R_0$ and resetting to $R_0$ with rate $r$. Clearly \cite{Redner_2001, Bray_2013} 
\begin{equation}
    F_r(R_0,t) = -\frac{\p Q_0(R_0,t)}{\p t}\,.
\end{equation}
Then the MFPT is just the first moment of the first passage probability density, and it is given by 
\begin{equation}
T_r(R_0)= \int_0^{+\infty}F_r(R_0,t) \,t\,\dd t=\int_0^{+\infty}\,  [-\partial_t\, Q_r(R_0, t)]\, t\, dt\,.
\end{equation}
Integrating by parts and using the initial condition given in Eq.~\eqref{eq:ic_backward_equation} gives
\begin{equation}
T_r(R_0) = \int_0^\infty Q_r(R_0,t)\, dt = \tilde{Q}_r(R_0,0).
\label{eq:MFPT1sm}
\end{equation}
Next, we use the relation between the survival probability with and without resetting (as given in Eq. (3) of the main text)
\begin{equation}
{\tilde Q}_r(R_0,s)= \frac{{\tilde Q}_0(R_0, r+s)}{1- r\, {\tilde Q}_0(R_0, r+s)}\, 
\label{renewal.SMm}
\end{equation}
in \eref{eq:MFPT1sm} with $s=0$. This gives
\begin{equation}
    T_r(R_0)= \frac{{\tilde Q}_0(R_0, r)}{1- r\, {\tilde Q}_0(R_0, r)}\, .
    \label{eq:MFPT2sm}
\end{equation}
On the other hand, the capture probability is given by (see Eq. (4) of the main text)
\begin{equation}
C(R_0)= \left[ \frac{1- (r+b)\, {\tilde Q}_0(R_0, r+b)}{1- r\, {\tilde Q}_0 (R_0, r+b)}\right]\, .
\label{capture.3msm}
\end{equation} 
 Now, it is easy to see that \eref{capture.3msm} can be expressed in terms of $T_r(R_0)$ in \eref{eq:MFPT2sm} as 
\begin{equation}
    C(R_0) = \frac{1}{1+b\, T_{r+b}(R_0)}\, .
\end{equation}
Therefore, maximizing the capture probability $C(R_0)$ with respect to the parameters $(r,\alpha)$ is equivalent to minimizing the MPFT $T_{r+b}(\vec{R}_0)$ with respect to $(r+b,\alpha)$. Recall from the main text that in $d$ dimensions (also see~\sref{sec:sp} for a derivation)
\begin{equation}\label{eq:surv_nr_m}
    \tilde Q_0(R_0, s) = \frac{1}{s}\left[1 -  \left(\frac{R_0}{\epsilon}\right)^{\frac{2-d}{2}} \frac{K_{(d-2)\mu}\left(2\mu \sqrt{s}R_0^{\frac{1}{2\mu}}\right)}{K_{(d-2)\mu}\left(2\mu \sqrt{s}\epsilon^{\frac{1}{2\mu}}\right)}\right]\, ,
\end{equation}
where $\mu = 1/(\alpha + 2)$, $K_\nu(x)$ is the modified Bessel function of the second kind, and $\epsilon$ is the size of the target.
Using~\eref{eq:surv_nr_m} in~\eref{eq:MFPT2sm}, we get
\begin{equation}\label{eq:MFPT_r_m}
    T_r(R_0) = \frac{1}{r}\left[ \left(\frac{R_0}{\epsilon}\right)^{(d-2)/2} \frac{K_{(d-2)\mu}\Big(2\mu\sqrt{r} \epsilon^{1/{2\mu}}
\Big)}{K_{(d-2)\mu}\Big(2\mu \sqrt{r}R_0^{1/{2\mu}}\Big)}-1\right]\, .
\end{equation}

\section{Universal power-law divergence of the optimal parameters as $R_0\to 1^+$}
\label{sec:pwb}

\subsection{$d=1$ case}
\label{sec:d1}
In (11) of the main text, we have reported the divergent behavior of  $\overline{r}$ and $\overline{\alpha}$ as the resetting distance $R_0$ from the target site approaches unity from above. Here, we provide a detailed derivation. Our results in Fig. 3 of the main text indicate that there is a divergence of  $\overline{r}$ and $\overline{\alpha}$ as $R_0\to 1^+$.  As mentioned in Sec.~\ref{sec:rel}, the problem of maximizing the capture probability $C(R_0)$ is equivalent to that of minimizing MPFT $T_{r+b}(R_0)$. It turns out that minimizing the MFPT is somehow easier than maximizing $C(R_0)$. In one dimension, using \eref{eq:MFPT_r_m}, the MFPT is given by 
\begin{equation}
\label{eq:1MFPT_resetting}
    T_{r+b}(R_0)=\frac{1}{r+b} \Bigg[\frac{  \Gamma \left(\mu\right)}{2 z^\mu \,K_{\mu}\left(2z\right)}-1\Bigg]\,,
\end{equation}
where we introduce a scaling variable 
\begin{equation}
\label{eq:scvar}
    z=\mu\sqrt{r+b}\, R_0^{1/(2\mu)}\,.
\end{equation}
\eref{eq:scvar} naturally suggests that one should keep the combination $\mu\sqrt{r+b}$ fixed when taking limit $\alpha\to \infty$, i.e., $\mu\to 0$ and $(r+b) \to \infty$. Since we assume $b$ is a constant $\sim O(1)$, $(r+b) \to \infty$ is equivalent to just $r \to \infty$.
In this limit, $R_0^{1/(2\mu)}\to \infty$ for $R_0>1$. We then use the asymptotic behavior 
\begin{equation}
\label{eq:asymK}
    K_\mu(2z) \sim \sqrt{\pi / (4z)}\,e^{-2z}\quad\text{as ~} z\to \infty\, .
\end{equation}
Taking the limit $\mu\to 0$, $r\to \infty$ keeping $\mu\sqrt{r}$ fixed and setting $R_0=1+\delta$ with $\delta >0$ small, in \eref{eq:1MFPT_resetting}, we get [using~\eref{eq:asymK}]
\begin{equation}
\label{eq:t1da}
    T_r(R_0) \approx \frac{1}{r} \left[\frac{r^{1/4}}{\sqrt{\mu\, \pi}} e^{(\frac{1}{4\mu}-\frac{1}{2})\delta}\, e^{2\mu\sqrt{r}\, e^{\delta/(2\mu)}}-1\right]. 
\end{equation}
We recall that, since $b\sim O(1)$ we have replaced the combination $(r+b)$ merely with $r$ everywhere.  Taking derivatives of~\eref{eq:t1da} with respect to $\mu$ and $r$ and equating them to zero yields
\begin{equation}
    2 \mu +4 \mu  \sqrt{r} \, e^{\frac{\delta }{2 \mu }} (\delta -2 \mu )+\delta =0
    \label{eq:tm0}
\end{equation}
and 
\begin{equation}
  e^{\frac{\delta}{4\mu}+2 \mu  \sqrt{r}\, e^{\frac{\delta }{2 \mu }}} \left(4 \mu  \sqrt{r} \,e^{\frac{\delta }{2 \mu }}-3\right)+4 \sqrt{\pi}\,   r^{-1/4} \sqrt{\mu } \, e^{\delta/2} =0
     \label{eq:tr0} 
\end{equation}
respectively. Solving \eref{eq:tm0} and \eref{eq:tr0} gives the optimal $r$ and $\mu$ as a function of $\delta$, where we recall that $\delta$ quantifies the proximity of $R_0$ to unity.  To this end, it is convenient to express the solution of the ratio $w=\mu/\delta$ and $\delta$. 
\Eref{eq:tm0} then becomes
\begin{equation}
\sqrt{r}= \frac{1}{\delta}\, G\left(\frac{\mu}{\delta}\right)\, ,\quad \text{where} \quad G(w) = \frac{1}{4\, w}\, \left(\frac{2w +1}{2w -1}\right)\, e^{-\frac{1 }{2w }}\, . 
\label{eq:rm0}
\end{equation}
We next substitute the expression of $\sqrt{r}$ from \eref{eq:rm0} in \eref{eq:tr0}, which gives
\begin{equation}
   f(w)=\delta\, e^{\delta/2}\quad\text{where}\quad  f(w) = \sqrt{\frac{2 w +1}{2w-1} }\,  \frac{(w-1)}{(2w-1)}\, \frac{1}{2\sqrt{\pi}\, w}\, e^{\frac{2w+1}{2(2w-1)}}\, .
    \label{eq:fw}
\end{equation}
From \eref{eq:fw}, when $\delta \to 0$, it is clear that  $w\to 1$. Expanding $f(w)$ around $w=1$ one gets to leading order, 
\begin{equation}
    f(w) \approx \frac{\sqrt{3}\, e^{3/2}}{2\sqrt{\pi}} (w-1).
    \label{eq:fw1}
\end{equation}
Therefore, inverting the relation $f(w)=\delta$ gives the optimal value of $w$ as
\begin{equation}
    w=1+\frac{2\sqrt{\pi}}{\sqrt{3}\,e^{3/2}}\, \delta +O(\delta^2).
\end{equation}

We recall that the optimal solutions for $r$ and $\alpha$ are denoted by $\overline{r}$ and $\overline{\alpha}$ respectively.
Since $\mu=\delta\, w$, the optimal exponent $\overline{\mu}=1/(\overline{\alpha}+2)$ is given by
\begin{equation}
    \frac{1}{\overline{\alpha}+2} = \delta+ O(\delta^2)
. 
\end{equation}
Consequently, to leading order in $\delta$, one gets
\begin{equation}
    \overline{\alpha} = \frac{1}{\delta} +O(1).
\end{equation}
Similarly, one can express the optimal value $\overline{r}$ in terms of $\delta$ as follows. Using the fact that when $\delta\to 0$, the ratio $w\to 1$. Hence, from \eref{eq:rm0}, one gets to leading order in $\delta$,
\begin{equation}
    \overline{r} \approx \frac{1}{\delta^2}\, G^2(1)  = \frac{9}{16 \,e \, \delta^2},
\end{equation}
where $e=\exp(1.0)$.
Hence, to summarize, to leading order in $\delta=R_0-1$, the optimal parameters $\{\overline{\alpha}, \overline{r}\}$ both diverge as 
\begin{equation}
    \overline{r}\approx \frac{9}{16 \,e \, (R_0 - 1)^2}\quad\text{and}\quad  \overline{\alpha} \approx \frac{1}{(R_0-1)}\,,
    \label{eq:ar-div}
\end{equation}
which is precisely (11) of the main text. 

\subsection{$d=2$ case}
\label{sec:d2}

In two dimensions, we again start with~\eref{eq:MFPT_r_m} and substitute $d=2$. For $\epsilon <1$, taking the limit $\mu\to 0$, $r\to \infty$ keeping $\mu\sqrt{r}$ fixed and setting $R_0=1+\delta$ with $\delta >0$ small, we get
\begin{equation}
    T_r(R_0) \approx \frac{1}{r}\left[-\left(\log \left(2 \mu  \sqrt{r}\right)+\frac{\ln \epsilon }{2 \mu }\right)\frac{2 \sqrt{\mu } \, r^{1/4}}{\sqrt{\pi}}\exp \left(\frac{\delta }{4 \mu }+2 \mu  \sqrt{r} \exp \left(\frac{\delta }{2 \mu }\right)\right) -1\right]. 
    \label{eq:tR02d}
\end{equation}
Taking a derivative of \eref{eq:tR02d} with respect to $\mu$ and setting it to zero gives
\begin{equation}
\label{eq:d2tm}
 \left[ 2 \mu   +4 \mu  \sqrt{r} e^{\frac{\delta }{2 \mu }} (\delta -2 \mu )+ \delta\right]\,  \ln \epsilon= 8 \mu ^2 
 +
 \ln \left(2 \mu  \sqrt{r}\right) \left[4 \mu ^2 - 2 \delta  \mu - 8 \mu  \sqrt{r}\, e^{\frac{\delta }{2 \mu }} (\delta -2 \mu )  \mu  \right]\, .
\end{equation}
Similarly, taking a derivative of \eref{eq:tR02d} with respect to $r$ and setting it to zero gives
\begin{equation}
    4\sqrt{\pi} \,r^{-1/4}\,\sqrt{\mu }= e^{\frac{\delta }{4 \mu }+2 \mu  \sqrt{r} e^{\frac{\delta }{2 \mu }}} \left[2 \mu  \left(\left[4 \mu  \sqrt{r} e^{\frac{\delta }{2 \mu }}-3\right] \ln \left(2 \mu  \sqrt{r}\right)+2\right)+ \left(4 \mu  \sqrt{r} e^{\frac{\delta }{2 \mu }}-3\right)\, \ln \epsilon \right].
    \label{eq:d2tr}
\end{equation}
Equations~\eqref{eq:d2tm} and \eqref{eq:d2tr} together determines $\mu$ and $r$ as a function of $\delta=R_0-1$. We are interested in their behavior as $\delta\to 0$. To this end, similar to the one-dimensional case in Sec.~\ref{sec:d1}, it is convenient to define the variable $w=\mu/\delta$. In the scaling limit $\delta\to 0$, $\mu\to 0$, and $r\to\infty$, keeping $\mu/\delta=w$ and $r\,\delta^2$ fixed, \eref{eq:d2tm} admits the scaling solution for $r$ as
\begin{equation}
\sqrt{r}= \frac{1}{\delta}\, G\left(\frac{\mu}{\delta}\right)\, ,\quad \text{where} \quad G(w) = \frac{1}{4\, w}\, \left(\frac{2w +1}{2w -1}\right)\, e^{-\frac{1 }{2w }}\, .  
\label{eq:rm02d}
\end{equation}
Interestingly, $G(z)$ in \eref{eq:rm02d}, is exactly the same scaling function that also appears in \eref{eq:rm0} in the $d=1$ case [Sec.~\ref{sec:d1}].  Using this scaling solution of $r$ in \eref{eq:d2tr} together with setting $\mu=w\,\delta$, gives an equation that determines $w$ in terms of $\delta$. This equation turns out to be
\begin{equation}
   f(w) =\frac{\delta}{(-\ln \epsilon)} \,  g(w)\, ,
  \label{eq:fw2d}
\end{equation}
where 
\begin{equation}
    g(w)=1- \frac{e^{\frac{2 w+1}{2 (2 w-1)}} \sqrt{2 w+1} }{2 \sqrt{\pi }\, w \,(2 w-1)^{3/2}}\, \left[2 w^2-1-2 (w-1) w \ln \left(\frac{2 w+1}{2 (2 w-1)}\right)\right]\, ,
    \label{eq:gw}
\end{equation}
and we recall that $f(w)$ in \eref{eq:fw2d} is the same function that appers in \eref{eq:fw}. From \eref{eq:fw2d}, it is clear that $w\to 1$ when $\delta\to 0$. Therefore, we expand \eref{eq:fw2d} around $w=1$ to the leading order,
\begin{equation}
    (w-1)  f'(1) =\frac{\delta}{(-\ln \epsilon)} \,  g(1) + O((w-1)^2)\,.  
\end{equation}
This gives
\begin{equation}
  w=1+ \frac{\delta}{(-\ln \epsilon)} \,  \frac{g(1)}{f'(1)}+O(\delta^2). 
  \label{eq:w2d}
\end{equation}

From \eref{eq:fw2d} and \eref{eq:gw}, we get 
\begin{equation}
    f'(1)= \frac{\sqrt{3}\, e^{3/2}}{2\sqrt{\pi}} \quad\text{and}\quad g(1)=1-\frac{\sqrt{3}\, e^{3/2}}{2\sqrt{\pi}}. 
    \label{eq:fg1}
\end{equation}
Finally, substituting \eref{eq:fg1} in \eref{eq:w2d} we get
\begin{equation}
    w=1+\left(1-\frac{2 \sqrt{\pi }}{\sqrt{3}\,e^{3/2}}\right)\frac{\delta}{ \ln \epsilon } +O(\delta^2).
\end{equation}

We recall that the optimal solutions for $r$ and $\alpha$ are denoted by $\overline{r}$ and $\overline{\alpha}$ respectively. Since $\mu=\delta\, w$, the optimal exponent $\overline{\mu}=1/(\overline{\alpha}+2)$ is given by
\begin{equation}
    \frac{1}{\overline{\alpha}+2} = \delta+ O(\delta^2)
. 
\end{equation}
Consequently, to leading order in $\delta$, one gets
\begin{equation}
    \overline{\alpha} = \frac{1}{\delta} +O(1).
\end{equation}
Similarly, one can express the optimal value $\overline{r}$ in terms of $\delta$ as follows. Note from~\eref{eq:w2d} that when $\delta\to 0$, the ratio $w$ approaches unity. Hence, from \eref{eq:rm02d}, one gets to leading order in $\delta$,
\begin{equation}
    \overline{r} \approx \frac{1}{\delta^2}\, G^2(1)  = \frac{9}{16 \,e \, \delta^2},
\end{equation}
where $e=\exp(1.0)$.
Hence, to summarize, to leading order in $\delta=R_0-1$, the optimal parameters $(\overline{r}, \overline{\alpha})$ both diverge as 
\begin{equation}
    \overline{\alpha} \approx \frac{1}{(R_0-1)} \quad\text{and}\quad \overline{r}\approx \frac{9}{16 \,e \, (R_0 - 1)^2}\,.
\end{equation}
Thus, in $d=2$, the leading order behavior of the optimal parameters $(\overline{r}, \overline{\alpha})$ as $R_0\to 1^+$ is independent of $\epsilon$. The $\epsilon$-dependence comes only in the sub-leading orders. This makes the leading order result universal, i.e., same as in $d=1$ [see Eq.~\eqref{eq:ar-div}].
\subsection{General $d>2$ case}
\label{sec:dg2}
For general $d> 2$, we again start with~\eref{eq:MFPT_r_m}. 
For $\epsilon <1$, taking the limit $\mu\to 0$, $r\to \infty$ keeping $\mu\sqrt{r}$ fixed and setting $R_0=1+\delta$ with $\delta >0$ small, we get
\begin{equation}
\label{eq:tg2}
    T_r(R_0) \approx \frac{1}{r}\left[\frac{r^{1/4}  }{| \gamma |\,\epsilon^{\frac{1}{2} (|\gamma | +\gamma )}\, \sqrt{\pi \, \mu}   }\, e^{  \left(\frac{1}{4 \mu }+ \frac{\gamma }{2}\right)\delta+2 \mu  \sqrt{r} e^{\frac{\delta }{2 \mu }}}-1 \right],\quad\text{where ~~} \gamma=d-2.
\end{equation}
Taking a derivative of~\eref{eq:tg2} with respect to $\mu$ and equating it to zero we get
\begin{equation}
    2 \mu  + 4 \mu  \sqrt{r} e^{\frac{\delta }{2 \mu }} (\delta -2 \mu ) +\delta=0.
    \label{eq:tm0gd}
\end{equation}
This is indeed exactly the same equation \eqref{eq:tm0} that we obtained for the $d=1$ case that admits the exact scaling solution for $r$ (for all $\delta$) as [see \eref{eq:rm0} for $d=1$, and \eref{eq:rm02d} for $d=2$ with $\delta\to 0$]
\begin{equation}
  \sqrt{r}= \frac{1}{\delta}\, G\left(\frac{\mu}{\delta}\right)\, ,\quad \text{where} \quad G(w) = \frac{1}{4\, w}\, \left(\frac{2w +1}{2w -1}\right)\, e^{-\frac{1 }{2w }}\, . 
\label{eq:rm0dd}  
\end{equation}
Note that this is independent of the target size $\epsilon$. Similar to both $d=1$ (Sec~\ref{sec:d1}) and $d=2$ (Sec~\ref{sec:d2}) case, taking a derivative of Eq. \eqref{eq:tg2} with respect to $r$,
\begin{equation}
    4 \sqrt{\pi}\,| \gamma |  \epsilon ^{\frac{| \gamma | +\gamma }{2}}\, r^{-1/4}\, \sqrt{\mu }\, e^{-\gamma/2} +\left(4 \mu  \sqrt{r} e^{\frac{\delta }{2 \mu }}-3\right) e^{\frac{\delta}{4\mu} +2 \mu  \sqrt{r} e^{\frac{\delta }{2 \mu }}}=0.
    \label{eq:tr0dd}
\end{equation}
We next substitute the expression of $\sqrt{r}$ from \eref{eq:rm0dd} in \eref{eq:tr0dd}, which gives
\begin{equation}
    f(w)=  | \gamma |  \epsilon ^{\frac{| \gamma | +\gamma }{2}}\, \delta\, e^{-\gamma\delta/2} , 
    \label{eq:fwd}
\end{equation}
where $f(w)$ is given in \eref{eq:fw}. Note that for $d=1$, we have $\gamma=-1$. Hence, \eref{eq:fwd} reproduces \eref{eq:fw} for $d=1$. For general $d>2$, from \eref{eq:fwd}, when $\delta\to 0$, we have $w\to 1$. Therefore, using the expansion of $f(w)$ near $w=1$ from \eref{eq:fw1}, we get 
\begin{equation}
    w=1+\frac{2\sqrt{\pi}}{\sqrt{3}}\, e^{-3/2}\, | \gamma |  \epsilon ^{\frac{| \gamma | +\gamma }{2}}\,\, \delta +O(\delta^2).
\end{equation}

We recall that the optimal solutions for $r$ and $\alpha$ are denoted by $\overline{r}$ and $\overline{\alpha}$ respectively. Since $\mu=\delta\, w$, the optimal exponent $\overline{\mu}=1/(\overline{\alpha}+2)$ is given by
\begin{equation}
    \frac{1}{\overline{\alpha}+2} = \delta+ O(\delta^2)
. 
\end{equation}
Consequently, to leading order in $\delta$, one gets
\begin{equation}
    \overline{\alpha} = \frac{1}{\delta} +O(1).
\end{equation}
Similarly, one can express the optimal value $\overline{r}$ in terms of $\delta$ as follows. Using the fact that when $\delta\to 0$, the ratio $w\to 1$. Hence, from \eref{eq:rm0dd}, one gets to leading order in $\delta$,
\begin{equation}
    \overline{r} \approx \frac{1}{\delta^2}\, G^2(1)  = \frac{9}{16 \,e \, \delta^2}~ ,
\end{equation}
where $e=\exp(1.0)$.
Hence, to summarize, to leading order in $\delta=R_0-1$, the optimal parameters $( \overline{r}, \overline{\alpha})$ both diverge as 
\begin{equation}
    \overline{r}\approx \frac{9}{16 \,e \, (R_0 - 1)^2} \quad\text{and}\quad \overline{\alpha} \approx \frac{1}{(R_0-1)}\,.
\end{equation}
Thus, as in  $d=2$, we find that the leading order behaviors of $\overline{r}$ and $\overline{\alpha}$ as $R_0\to 1^+$ are independent of $\epsilon$. Moreover, they are exactly the same for all $d>2$ and $d=1,2$. Hence, the leading order behaviors of the optimal parameters $\overline{r}$ and $\overline{\alpha}$ as $R_0\to 1^+$ are universal in the sense that they do not depend on the dimension $d$.

\section{Capture probability in the presence of a finite lifetime resetting strategy}
\label{Model_cp}
As reported in the main text in (4), the capture probability, $C(\vec{R}_0)\equiv C(R_0)$, is given by 
\begin{equation}
C(R_0)= \left[ \frac{1- (r+b)\, {\tilde Q}_0(R_0, r+b)}{1- r\, {\tilde Q}_0 (R_0, r+b)}\right]\, ,
\label{capture.3}
\end{equation}
where the Laplace transform of the survival probability is given in~\eref{eq:surv_nr_supp}.
This gives the exact expression of the capture probability within this strategy
\begin{equation}
 C(R_0)=\frac{(b+r)}{r + b\,  \left(\frac{R_0}{\epsilon}\right)^{(d-2)/2}\,
\frac{ K_{(d-2)\mu}\left(2\mu\, (r+b)^{1/2}\, \epsilon^{1/{2\mu}}\right)}
{ K_{(d-2)\mu}\left(2\mu\, (r+b)^{1/2}\, R_0^{1/{2\mu}}\right)} }.
\label{capture_final}
\end{equation}

Our goal now is to maximize the function $C(R_0)$ with respect to the two
parameters $(r,\alpha)$ for fixed $R_0$, $b$, and $\epsilon$. It turns out that it is easier to work with the parameters $(r, \mu)$ [where $\mu=1/(\alpha + 2)$] rather than with $(r, \alpha)$.
The optimal parameters $(\overline{r},\overline{\mu})$ that maximizes $C(R_0)$ can be obtained by solving simultaneously the pair of equations 
\begin{equation}
        \p_r C(R_0)\Big|_{(\overline{r},\overline{\mu})} = 0\quad\text{and}\quad
        \p_\mu C(R_0)\Big|_{(\overline{r},\overline{\mu})}  = 0\, .
        \label{eq:pair}
\end{equation}
Differentiating Eq.~\eqref{capture.3} with respect to $r$,  we get
\begin{equation}
    \p_r C(R_0) = - b \frac{\p_r\tilde  Q_0(R_0,r+b)+Q_0^2(R_0,r+b)}{[1-r Q_0(R_0,r+b)]^2}\,.
    \label{eq:dC}
\end{equation}
Therefore, using~\eref{eq:dC} in the first condition in~\eref{eq:pair} gives
\begin{equation}
     \left[\p_r\tilde Q_0(R_0,r+b)+\tilde Q_0(R_0,r+b)^2\right]\Big|_{(\overline{r},\overline{\mu})} =0\,.
     \label{eq:cond1}
\end{equation}
Similarly, differentiating~\eref{capture.3} with respect to $\mu$ yields
\begin{equation}
    \p_\mu C(R_0) =- b \frac{\p_\mu\tilde  Q_0(R_0,r+b)}{[1-r Q_0(R_0,r+b)]^2}\,.
    \label{eq:dmC}
\end{equation}
Using~\eref{eq:dmC} in the second condition in~\eref{eq:pair} gives
\begin{equation}
 \p_\mu\tilde  Q_0(R_0,r+b)\Big|_{(\overline{r},\overline{\mu})} = 0\, . 
 \label{eq:cond2}
\end{equation}
Equations~\eqref{eq:cond1} and \eqref{eq:cond2} together determines the optimal pair of the parameters $(\overline{r},\overline{\mu})$, or equivalently, $(\overline{r},\overline{\alpha})$. In the following subsections, we analyze these conditions explicitly in different dimensions.    

For simplicity, we consider the case of one dimension. For $d=1$, one can take the limit of the target size going to zero. In the limit $\epsilon\to 0$ and $d=1$,~\eref{eq:surv_nr_supp} becomes
\begin{equation}\label{eq:SP_d1_capt}
    \tilde Q_0(R_0,s) = \frac{1}{s}\left[1 - \frac{2z^\mu K_\mu(2z)}{\Gamma(\mu)}\right]\quad \text{where}\quad z = \mu \sqrt{s}\,R_0^{\frac{1}{2\mu}}\,.
\end{equation}
Using \eref{eq:SP_d1_capt} in Eq.~\eqref{eq:cond1} yields,
\begin{equation}
   2 \bar{z}^{\overline{\mu} } K_{\overline{\mu} }^2 (2 \bar z)+\bar{z} \Gamma (\overline{\mu} ) K_{\overline{\mu} -1}(2
  \bar z)-\Gamma (\overline{\mu} ) K_{\overline{\mu} }(2 \bar z)=0\,\quad\text{where}\quad \bar{z} = \overline{\mu}\, \sqrt{\overline{r}+b}\, R_0^{\frac{1}{2\overline{\mu}}}\, .
  \label{eq:cond1d1}
\end{equation}
Similarly, using \eref{eq:SP_d1_capt} in Eq.~\eqref{eq:cond1} gives
\begin{equation}
 \overline{\mu}^2 K_{\overline{\mu}}^{(1,0)}(2 \bar{z})+\bar{z} \left[\ln R_0-2 \overline{\mu} \right] K_{\overline{\mu} -1}(2 \bar{z})+\overline{\mu} ^2 K_{\overline{\mu} }(2 \bar{z}) [\ln \bar{z}-\psi ^{(0)}(\overline{\mu} )]=0\, ,
\label{eq:cond1d2}
\end{equation}
where $K_{\overline{\mu}}^{(1,0)}(2 \bar{z}) =\p_{\overline{\mu}}K_{\overline{\mu} }(2 \bar{z}) $ and $\psi ^{(0)}(\overline{\mu} )$ is the polygamma function. Solving Eqs.~\eqref{eq:cond1d1} and \eqref{eq:cond1d2} together gives $\overline{\mu}$ and $\bar{z}$ for a given $R_0$, which in turn, determines the optimal parameters $\overline{r}(R_0)$ and $\overline{\mu}(R_0)$.   

Given $\overline{\mu}$,   Eq.~\eqref{eq:cond1d1} determines $\bar{z}$ uniquely. Therefore, formally we can express the optimal $\overline{r}$, using the last expression in Eq.~\eqref{eq:cond1d1}, as
\begin{equation}
    \overline{r} = \left[\frac{\bar z(\overline{\mu})}{\mu}\right]^2\frac{1}{R_0^{1/\mu}} - b\,.
    \label{eq:ropt}
\end{equation}
Since the resetting rate in Eq.~\eqref{eq:ropt} must be non-negative, there exists a certain $R_{0}^*$ such that
\begin{equation}\label{eq:r_star_1d_capture}
    \overline{r} (R_0) = \begin{cases}
        0 & R_0>R^*_{0}(\overline{\mu})\\[3mm]
        \left(\frac{\bar z(\overline{\mu})}{\overline{\mu}}\right)^2\frac{1}{R_0^{1/\overline{\mu}}} - b & R_0\leq R^*_{0}(\overline{\mu})
    \end{cases}
\end{equation}
where
\begin{equation}
    R^*_{0}(\overline{\mu}) = \left(\frac{\bar z(\overline{\mu})}{\overline{\mu}}\right)^{2\,\overline{\mu}}\frac{1}{b^{\overline{\mu}}}\,.
\label{eq:R0*}    
\end{equation}

Indeed, by solving Eqs.~\eqref{eq:cond1d1} and \eqref{eq:cond1d2} numerically, as shown in Fig. 3 of the main text, we find that $\overline{r}=0$ and $\overline{\mu}=1/2$ for large $R_0$. Therefore, solving Eq.~\eqref{eq:cond1d1} numerically with $\overline{\mu}=1/2$, we get $\bar{z}(1/2)=0.79681\dots$. Consequently, from Eq.~\eqref{eq:R0*} we get
\begin{equation}
    R_{02}^*\equiv R_0^*(\overline{\mu}=1/2) = \frac{2\bar{z}(1/2)}{\sqrt{b}}= \frac{1.5936\dots}{\sqrt{b}}
    \label{eq:R02s}
\end{equation}

From Fig. 3~(a) of the main text, there exists another critical point $R_{01}^* < R_{02}^*$ such that $\overline{\mu}$ still remains $1/2$ in-between. In this range, $\overline{r}$  is given by
\begin{equation}
    \overline{r}= \frac{[2\bar{z}(1/2)]^2}{\sqrt{R_0}}-b\quad\text{where}\quad R_{01}^* < R_0 < R_{02}^*\, .
    \label{eq:opt2h}
\end{equation}

At $R_0=R_{01}^*$, both $\overline{\alpha}$ and $\overline{r}$ undergo a phase transition. The critical point $R_{01}^*$ can be obtained by solving Eq.~\eqref{eq:cond1d2} with $\overline{\mu}=1/2$ and $\bar{z} = \bar{z}(1/2)=0.79681\dots$. This yields
\begin{equation}
    R_{01}^*= \exp\left\{-\frac{\gamma_E}{4\bar{z}(1/2)} + 1 - \frac{1}{4\bar{z}(1/2)}\ln [4 \bar{z}(1/2)] - \frac{e^{2 \bar{z}(1/2)}}{\sqrt{4\pi \bar{z}(1/2)}}\,K_{1/2}^{(1,0)}\big(2\bar{z}(1/2)\big)\right\} =1.4578\dots
    \label{eq:R01s}
\end{equation}
where $K_{1/2}^{(1,0)}(x) =\p_\mu K_{\mu}(x)|_{\mu = 1/2}$, which has a nice expression in terms of the exponential integral 
\begin{equation}\label{eq:bessel_der_exp_int}
    K_{1/2}^{(1,0)}(x) = \sqrt{\frac{\pi}{2x}}E_1(2x) e^x, \quad\text{where}\quad
       E_1(x) = \int_{-\infty}^x\frac{e^t}{t} \,dt
\end{equation}
is the exponential integral.

From Eqs~\eqref{eq:R02s} and \eqref{eq:R01s}, the condition $R_{01}^* < R_{02}^*$ demands that
\begin{equation}
    b < b^* \quad\text{where}\quad b^*= 1.19494\dots
\end{equation}

To summarize, for $b<b^*$, there are two critical points $R_{01}^*$ and $R_{02}^*$ such that 
\begin{alignat}{3}
        &R_0 > R_{02}^*~: &\qquad& \overline{\alpha}=0,  &\quad& \overline{r}=0\\
        &R_{01}^* < R_0 < R_{02}^* ~: && \overline{\alpha}=0, && \overline{r} > 0\\
        &1 < R_0 < R_{01}^* ~: && \overline{\alpha}>0, && \overline{r}>0\\
        &\epsilon < R_0 <1 ~:  && \overline{\alpha}\to\infty, && \overline{r}\to\infty
\end{alignat}

On the other hand, for $b>b^*$, a different picture emerges. There are still two critical points $R_{03}^*$ and $R_{04}^*$. For $R_0 > R_{04}^*$, the optimal parameters $\overline{r}$ and $\overline{\alpha}$ are still zero, as seen in Fig. 3~(b) of the main text. In this case, $\overline{\alpha}$ undergoes a phase transition at $R_{04}^*$, while $\overline{r}$ still remains zero across $R_{04}^*$. Hence, the critical value $R_{04}^*$ can be obtained by setting $\overline{\mu}=1/2$ and $\bar{z}=\frac{1}{2}\sqrt{b}\,R_{04}^* $ in Eq.~\eqref{eq:cond1d2}. This yields,
\begin{equation}
 \sqrt{\frac{2}{\pi}}\, b^{1/4}\sqrt{R_0}\, K_{1/2}^{(1,0)}\left(\sqrt{b} \,R_0\right)+e^{-\sqrt{b} \,R_0} \left[-2 \sqrt{b} \,R_0+\ln \left(2 \sqrt{b}\, R_0\right)+2 \sqrt{b} \,R_0 \ln \left(R_0\right)+\gamma_E \right] =0\, .
 \label{eq:R04}
\end{equation}
Equation~\eqref{eq:R04} determines $R_{04}^*$ as a function of $b$ for $b>b^*$. For example, for $b=2$, numerical solution of Eq.~\eqref{eq:R04} yields, $R_{04}^*=1.6661\dots$. The optimal value of $\overline{\alpha}$, or equivalently $\overline{\mu}$ for $R_{03}^* < R_0< R_{04}^*$, can be obtained by solving Eq.~\eqref{eq:cond1d2} as a function of $b$ and $R_0$, by setting $\bar{z}= \mu\sqrt{b}\, R_0^{\frac{1}{2\mu}}$. Fig. 3~(b) of the main text shows the optimal $\overline{\alpha}$
as a function of $R_0$. When $R_0$ hits $R_{03}^*$ from above, both $\overline{\alpha}$ and $\overline{r}$ undergo another phase transition. Since $\overline{r}$ undergoes a zero to non-zero transition at $R_{03}^*$, from Eq.~\eqref{eq:R0*}, $R_{03}^*$ should self-consistently satisfy the relation
\begin{equation}
    R^*_{03} = \left(\frac{\bar z(\mu^*)}{\mu^*}\right)^{2\mu^*}\frac{1}{b^{\mu^*}}\, \quad\text{where}\quad \mu^* =\overline{\mu}(R_{03}^*).
\label{eq:R03}    
\end{equation}
This can indeed be checked at the critical point $R_{03}^*$ for $b=2$ in Fig. 3~(b) of the main text. 

To summarize, for $b>b^*$, we have
\begin{alignat}{3}
        &R_0 > R_{04}^*~: &\qquad& \overline{\alpha}=0,  &\quad& \overline{r}=0\\
        &R_{03}^* < R_0 < R_{04}^* ~: && \overline{\alpha}>0, && \overline{r} = 0\\
        &1 < R_0 < R_{03}^* ~: && \overline{\alpha}>0, && \overline{r}>0\\
        &\epsilon < R_0 <1  ~:&& \overline{\alpha}\to\infty, && \overline{r}\to\infty
\end{alignat}

A similar analysis can be performed in higher dimensions, but there is an additional parameter $\epsilon$.

\section{Optimal and non-optimal trajectories in $d=2$}
\label{sec:supp:traj}

\begin{figure}   
\includegraphics[scale=0.39]{main_figure_trajectory1.pdf}\quad\quad\quad\quad\quad\quad
\includegraphics[scale=0.39]
{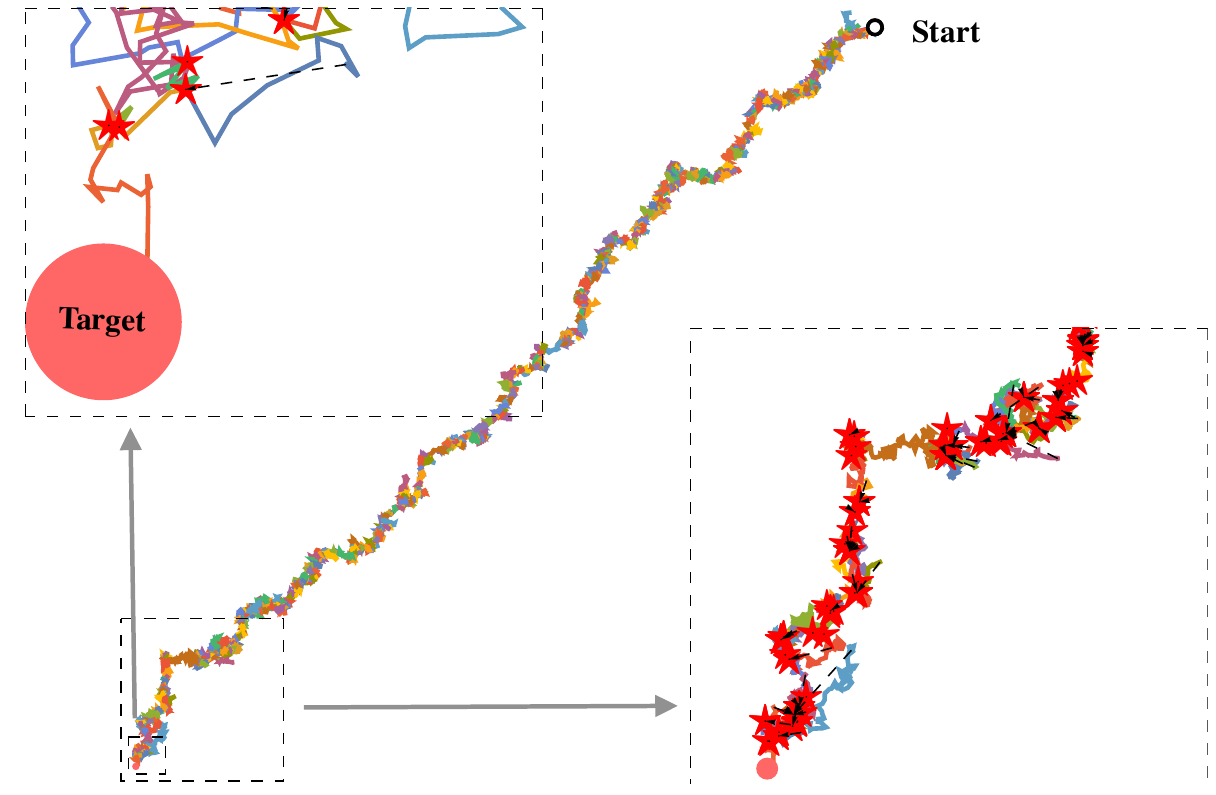}
    \caption{(Left Panel) An optimal trajectory in $d=2$, for fixed initial position $\vec{R}_0 = (100, 100)$, $b= 0.2$ and target size $\epsilon=0.5$. This trajectory is optimal in the sense that at each time interval $\tau_i$ of the search process we set the two parameters ($r,\alpha$) to be the optimal ones (those maximizing the capture probability in that interval), i.e., $(r, \alpha) = (\bar r, \bar \alpha)$. 
        (Right Panel) A \emph{sub-optimal} trajectory generated by the \emph{proxitaxis} search strategy in $d=2$ with the same values of $\bar{R}_0,b$ and $\epsilon$ as in the left panel. The trajectory is sub-optimal because in each time interval $\tau_i$ of the process the parameters $(r,\alpha)$ are not chosen to be the optimal ones.  
        Rather, the values of $(r,\alpha)$ are fixed for the entire process to be $r=0.05$ and $\alpha = 0.1$ until the detection is achieved. 
        The single target is shown by the red disk and the searcher's initial location is indicated by the black circle. The central figure shows the trajectory over a large length and time scale, illustrating that the strategy guides the searcher almost deterministically towards the target. However, on smaller scales, one sees fluctuations. For example, the top box zooms the trajectory close to the target, where the searcher becomes more active. In contrast, the right box zooms the trajectory at some intermediate scales where it is less active, i.e., more sluggish. See the main text for the details of the strategy and Sec.~\ref{sec:supp:traj} for the details of numerical implementation. 
   } 
    \label{fig:traj_supp}
\end{figure}

In this section, for fixed $R_0$ and $b$, we present two trajectories for $d=2$ (a) corresponding to the optimal choice of the parameters $(\bar{r}_i,\bar{\alpha}_i)$ in each time interval $i$ between two successive inspections and (b) where $(r,\alpha)$ are fixed for all intervals and are therefore not optimal. These are shown respectively in the left and the right panels of Fig.~\ref{fig:traj_supp}.

We now explain briefly how Fig.~1 of the main text and Fig. \ref{fig:traj_supp} were obtained. To simulate a trajectory of the \emph{proxitaxis} strategy using a computer, one needs to implement the following algorithm. The input parameters are the value of $b>0$, the starting position of the searcher $\vec{R}_0^{(1)}$, the dimension $d$, and the size of the target $\epsilon>0$ (for $d=1$ one can also take $\epsilon=0$). For these initial values of $b$ and $\vec{R}_0^{(1)}$ we optimize numerically the capture probability Eq. \eqref{capture_final} and calculate $(\bar r_1, \bar \alpha_1)$ for this time interval. Then we draw a random variable $\tau_1$ from an exponential distribution with mean $b^{-1}$ and evolve the position of the searcher according to the dynamics described by the equations in Eq.~\eqref{eq:langeving_d_dim} with initial condition $\vec{R}_0^{(1)}$ and up to time $\tau_1$. After this time $\tau_1$ one sets checks the distance from the target: if $\|\vec{R}(\tau_1)\|<\|\vec{R}_0^{(1)}\|$ then the new initial position is $\vec{R}^{(2)}_0=\vec{R}(\tau_1)$; otherwise $\vec{R}^{(2)}_0=\vec{R}^{(1)}_0$. At this point one recomputes the optimal parameters $(\bar r_2, \bar \alpha_2)$ for the new initial condition $\vec{R}_0^{(2)}$. Then one draws another random time $\tau_2$ from the same exponential distribution and follows exactly the same steps described before. This results in a sequence of optimized values $(\bar r_i, \bar \alpha_i)$ for $i=1,2,\dots$ until the target is reached. The strategy is efficient because the probability of capture is optimized within each epoch $\tau_i$.

\section{Optimization of the capture probability in two dimensions} 
\begin{figure*}[h]
    \centering
    \includegraphics[width=.8\textwidth]{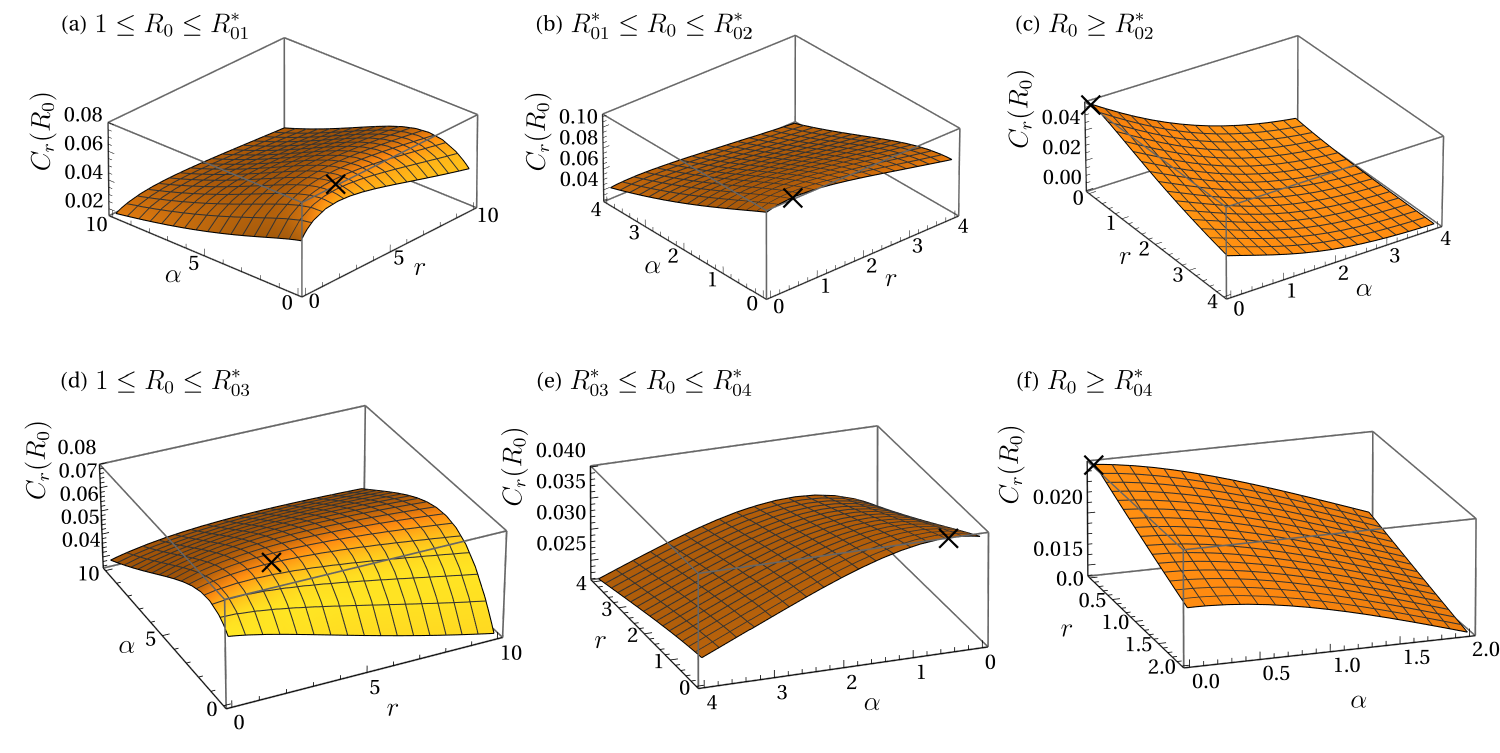}
     \centering
    \caption{
    Plots of the capture probability $C(R_0)$ in~\eref{eq:capture2d} as a function of $(r,\alpha)$ for different values of $R_0$ in two dimensions $d=2$ and target size $\epsilon=0.2$. The cross denotes the maximum  of $C(R_0)$ at the optimal values $(\overline{\alpha}, \overline{r})$. The top panels are for $b=1<b^*$ while the bottom panels are for $b=4>b^*$. (a) $R_0<R_{01}^*$ (b) $R_{01}^* <R_0<R_{02}^*$ (c) $R_0>R_{02}^*$ (d) $R_0<R_{03}^*$ (e) $R_{03}^* <R_0<R_{04}^*$ (f) $R_0>R_{04}^*$. Note that in general $R_{0i}^*$ for $i=2,3,4$ are non-trivial functions of $b$ (see~\fref{fig:optimal-alpha-r_EM1} for exact values corresponding to $b=1$ and $b=4$).}
    \label{fig:optimalCP_EM1}
\end{figure*}

\begin{figure}[h]
    \centering
    \includegraphics[width=0.8\linewidth]{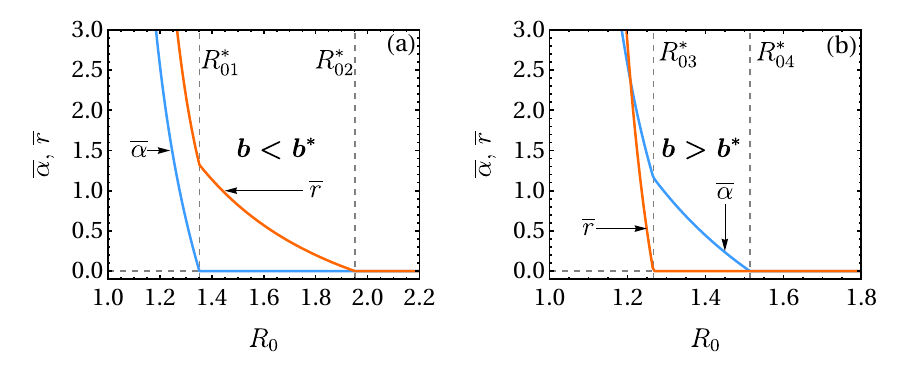}
    \caption{Plot of $\overline{r}$ and $\overline{\alpha}$ , as a function of $R_0$ for (a) $b<b^*$ and (b) $b>b^*$, in two dimensions $d=2$ and target size $\epsilon=0.2$, obtained by maximizing~\eref{eq:capture2d}. 
    For both cases, $\overline{r}$ and $\overline{\alpha}$ are nonanalytic at some critical values of $R_0$ marked by the vertical dash lines. For panel (a) we choose $b=1<b^*$, for which the critical values are at $R_{01}^*\approx 1.3524$ and $R_{02}^*\approx 1.9518$.
    For (b), we choose
    $b=4 > b^*$, for which the critical values of $R_0$ are at $R_{03}^*\approx 1.2663$ and $R_{04}^*\approx 1.5142$.
    }
    \label{fig:optimal-alpha-r_EM1}
\end{figure}

We discuss the capture probability in the case of $d=2$. In this case, the capture probability in $d$ dimensions [Eq.~\eqref{capture_final}]
becomes
\begin{equation}
\label{eq:capture2d}
    C(R_0)=\frac{(b+r)}{r + b\,
\frac{ K_{0}\left(2\mu\, \sqrt{r+b}\, \epsilon^{1/{2\mu}}\right)}
{ K_{0}\left(2\mu\, \sqrt{r+b}\, R_0^{1/{2\mu}}\right)} }\, ,
\end{equation}
where we recall that $\mu=1/(2+\alpha)$ and $\epsilon$ is the size of the object. 
We plot $C(R_0)$ given in~\eref{eq:capture2d} in the $(r,\alpha)$ plane in~\fref{fig:optimalCP_EM1} and see that for any fixed $R_0$, there is indeed a unique global maximum at $(\overline{r}, \overline{\alpha})$. These optimal parameters undergo a phase transition at two critical values of $R_0$ as demonstrated in~\fref{fig:optimal-alpha-r_EM1}.

\section{Optimization of the capture probability in three dimensions} 
\begin{figure}[h]
    \centering
    \includegraphics[width=.8\textwidth]{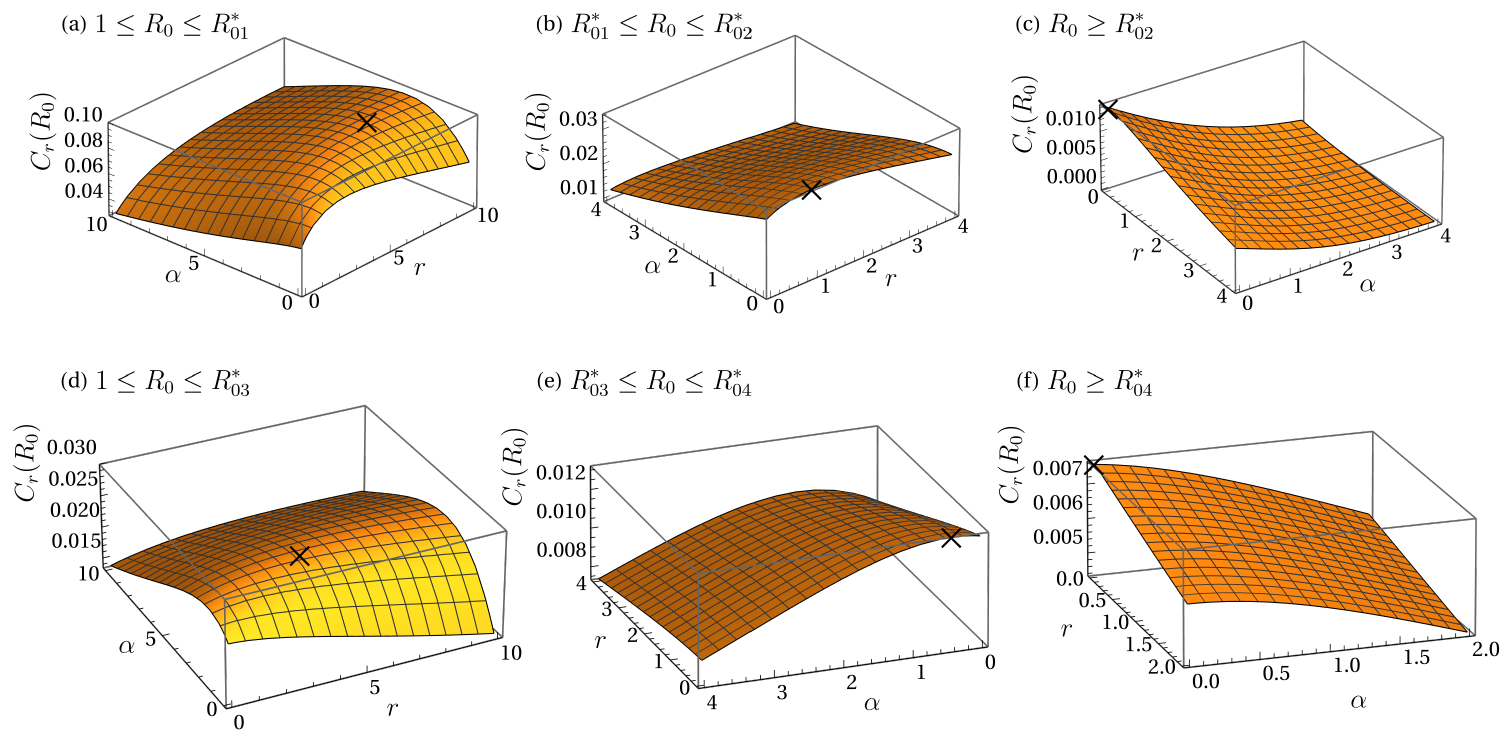}
     \centering
    \caption{Plots of the capture probability $C(R_0)$ in~\eref{capture_3D} as a function of $(r,\alpha)$ for different values of $R_0$ in two dimensions $d=3$ and target size $\epsilon=0.2$. The cross denotes the maximum  of $C(R_0)$ at the optimal values $(\overline{\alpha}, \overline{r})$. The top panels are for $b=1<b^*$ while the bottom panels are for $b=4>b^*$. (a) $R_0<R_{01}^*$ (b) $R_{01}^* <R_0<R_{02}^*$ (c) $R_0>R_{02}^*$ (d) $R_0<R_{03}^*$ (e) $R_{03}^* <R_0<R_{04}^*$ (f) $R_0>R_{04}^*$. Note that in general $R_{0i}^*$ for $i=2,3,4$ are non-trivial functions of $b$ (see~\fref{fig:optimal-alpha-r_EM1} for exact values corresponding to $b=1$ and $b=4$).}
    \label{fig:optimalCP_EM2}
\end{figure}

\begin{figure}[h]
    \centering
    \includegraphics[width=0.8\linewidth]{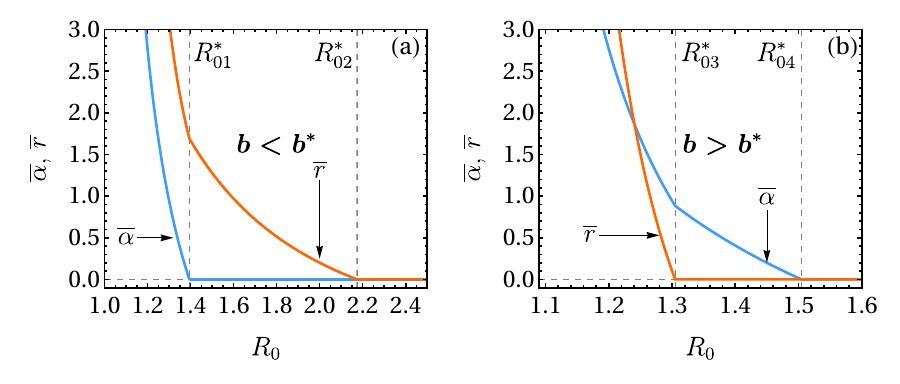}
    \caption{
    Plot of $\overline{r}$ and $\overline{\alpha}$, as a function of $R_0$ for (a) $b<b^*$ and (b) $b>b^*$, in three dimensions $d=3$ and target size $\epsilon=0.2$, obtained by maximizing~\eref{capture_3D}. 
    For both cases, $\overline{r}$ and $\overline{\alpha}$ are nonanalytic at some critical values of $R_0$  marked by the vertical dash lines. For panel (a) we choose $b=1<b^*$, for which the critical values are at $R_{01}^*\approx 1.3943$ and $R_{02}^*\approx 2.1743$.
    For (b), we choose
    $b=4 > b^*$, for which the critical values of $R_0$ are at $R_{03}^*\approx 1.3052$ and $R_{04}^*\approx 1.5049$.
    }
    \label{fig:optimal-alpha-r_EM2}
\end{figure}

For the $d=3$ case, the capture probability given in ~\eref{capture_final} becomes
\begin{equation}
 C(R_0)=\frac{(b+r)}{r + b\,  \left(\frac{R_0}{\epsilon}\right)^{\frac{1}{2}}\,
\frac{ K_{\mu}\left(2\mu\, \sqrt{r+b}\, \epsilon^{1/{2\mu}}\right)}
{ K_{\mu}\left(2\mu\, \sqrt{r+b}\, R_0^{1/{2\mu}}\right)} }\, ,
\label{capture_3D}
\end{equation}
where we recall that $\mu=1/(2+\alpha)$ and $\epsilon$ is the size of the target. 
Here again, 
we plot $C(R_0)$ given in~\eref{capture_3D} in the $(r,\alpha)$ plane in~\fref{fig:optimalCP_EM2} and see that for any fixed $R_0$, there is indeed a unique global maximum at $(\overline{r}, \overline{\alpha})$. Similar to the one and two-dimensional cases, these optimal parameters undergo a phase transition at two critical values of $R_0$ as demonstrated in~\fref{fig:optimal-alpha-r_EM2}.

\clearpage
\bibliography{supp_mat_references}